\documentclass[12pt]{iopart}

\usepackage{amsmath}
\usepackage{amssymb}
\usepackage{dsfont}
\usepackage{url}
\usepackage{graphicx}
\usepackage{multirow}  
\usepackage{lscape}
\usepackage{siunitx}
\usepackage{caption}
\usepackage{subcaption}
\usepackage{ulem}
\usepackage{xcolor}

\usepackage{lineno}
\usepackage[T1]{fontenc}

\begin{document}
\title[Optimizing searches for GW bursts using cWB--2G]{Optimizing searches for gravitational wave bursts using coherent WaveBurst--2G.}



\author{
A.~Martini$^{1,2}$,
A.~Miani$^{1,2}$,
M.~Drago$^{3,4}$,
C.~Lazzaro$^{5,6}$,
F.~Salemi$^{3,4}$,
S.~Bini$^{7}$,
O.~Freitas$^{8,9}$,
E.~Milotti$^{10,11}$,
G.~Principe$^{10,11}$,
S.~Tiwari$^{12}$,
A.~Trovato$^{10,11}$,
G.~Vedovato$^{13}$,
Y.~Xu$^{14}$,
and G.~A.~Prodi$^{1,2}$
}

\address{$^{1}$ Department of Physics, University of Trento, Trento, Italy}
\address{$^{2}$ INFN-TIFPA, Trento Institute for Fundamental Physics and Applications, Trento, Italy}
\address{$^{3}$ Department of Physics, Sapienza University of Rome, Rome, Italy}
\address{$^{4}$ INFN, Sezione di Roma, Rome, Italy}
\address{$^{5}$ Department of Physics, University of Cagliari, Cagliari, Italy}
\address{$^{6}$ INFN, Sezione di Cagliari, Cagliari, Italy}

\address{$^{7}$ LIGO Laboratory, California Institute of Technology, Pasadena, CA 91125, USA}
\address{$^{8}$ Department of
  Astronomy and Astrophysics, University of Valencia,  Valencia, Spain}
  \address{$^{9}$ CF-UM-UP, University of Minho, Braga, Portugal}
\address{$^{10}$ Department of Physics, University of Trieste, Trieste, Italy}
\address{$^{11}$ INFN, Sezione di Trieste, Trieste, Italy}

\address{$^{12}$ Department of Physics, University of Zurich, Zurich, Switzerland}
\address{$^{13}$ INFN, Sezione di Padova, Padova, Italy}
\address{$^{14}$ Universitat de les Illes Balears, Palma, Spain}

\ead{alessandro.martini-1@unitn.it}


\begin{abstract}
The most general searches for gravitational wave transients (GWTs) rely on data analysis methods that do not assume prior knowledge of the signal’s waveform, direction, and arrival time on Earth.
These searches provide data-driven signal reconstructions that are crucial both for testing available emission models and for discovering yet-to-be-uncovered sources. 
Here, we discuss the progresses in detection performances of the coherent WaveBurst second generation pipeline (cWB-2G), which is highly adaptable to minimally--modeled as well as model--informed searches for GWTs. Several search configurations for GWTs are examined using approximately 14.8 days of observation time from the third observing run by LIGO-Virgo-KAGRA (LVK).
Recent enhancements include a ranking statistic fully based on multivariate classification with eXtreme Gradient Boosting, a thorough validation of the accuracy of the statistical significance of GWT candidates, and the measurement of the correlations of false alarms and simulated detections between different concurrent searches. 
For the first time, we provide a comprehensive comparison of cWB-2G performances on data from networks made of two and three detectors, and we demonstrate the advantage of combining concurrent  searches for GWTs of generic morphology in a global observatory. 
This work offers essential insights for assessing our data analysis strategies in the ongoing and future LVK searches for generic GWTs.
\end{abstract}

\section{Introduction}
\label{sec:intro}
The observatories of the LIGO-Virgo-KAGRA (LVK) collaboration \cite{aLIGO, aVIRGO} are currently taking data in the fourth observing run (O4), with a catalog of more than two hundred gravitational wave transient (GWT) detections already published
 \cite{GWTC-1,GWTC-2,GWTC-3, GWTC-4results, GWOSC}. 
All detected sources so far belong to the Compact Binary Coalescence (CBC) class, for which the theory of General Relativity (GR) provides detailed models which are suitable for matched-filter searches and for testing  fundamental physics \cite{TGR1,TGR2,TGR3}.
In the future, we expect to observe other sources of GWTs for which there are no models as precise as those for CBCs, such as core--collapse supernovae (CCSN) \cite{Abdikamalov2022}, neutron star (NS) excitations \cite{GWfromMagnetars2025, starquakes2022}, nonlinear memory effects \cite{PhysRevD.101.104041}, and accretion disk instabilities \cite{collapsardiskGottlieb2024}.
In addition, in this pioneering phase of gravitational wave astronomy, we need to perform searches for GWTs from unexpected sources. 
For all these GWT sources, LVK  employs approaches  which are either fully or partially agnostic to signal morphologies. 
The broadest searches for GWTs --- called ``all-sky burst searches'' --- are designed to detect generic signals from any direction in the sky and at all times. This is a highly challenging task, and the ability to distinguish genuine GWTs from noise outliers is paramount.
These all-sky searches are then specialized to look for short--duration burst GWTs (up to a few seconds in the sensitive frequency band \cite{allsky1,allsky2,allsky3, allsky4}), and long--duration burst GWTs (up to $\sim 10^{3}$ s  \cite{long1,long2,long3, long4}).

This work reports recent developments and tests carried out on one flavor of the data analysis pipeline coherent WaveBurst (cWB) \cite{cwb2005, cwb2008, cwb2016}, that contributed to all LIGO-Virgo and LVK minimally--modeled all--sky searches for GWTs since the first GW detection GW150914 \cite{gw150914_discovery}, and played a crucial role in the detection of GW190521 \cite{gw190521_discovery} and GW231123 \cite{GW231123}. 
cWB carries out both low--latency searches and more thorough analyses on archived data,  
and it is capable of covering the entire sensitive frequency band.
Several versions of cWB are actively being developed, known as cWB second generation (cWB-2G \cite{cwb2016,cwbsftX, cwb2023}), cWB cross power (cWB-XP \cite{cwbxp2022,cwbxp2025}) and cWB with Gaussian Mixture Model post-processing (cWB-GMM \cite{cwbGMM2022,cwbGMM2024,cwbGMM2025}).



Here, we present the latest methodological advancements of cWB-2G \cite{cWB-6.4.6}, which are exploited in the analysis of the ongoing fourth LVK observing run. Several search configurations for GWTs are tested on data from the LVK observing run O3, the most recent 
publicly released data set including both LIGO \cite{aLIGO} and Virgo \cite{aVIRGO} observations.
In particular, we select 14.8 days of concurrent observation time by LIGO--Hanford (H), LIGO--Livingston (L) and Virgo (V).
Recent enhancements with respect to what was previously reported in \cite{Marek:2023} include the deployment of a ranking statistic fully based on machine learning methods,  a thorough validation of the accuracy of the statistical significance of GWT candidates, and the measurement of the correlations of false alarms and simulated detections between different concurrent searches. 
For the first time, thanks to our comprehensive comparison of search performances when using two and three detectors' data, we are able to gauge the advantages and drawbacks of combining concurrent  searches of GWTs of generic morphology in a global observatory of non-aligned detectors.

The paper is organized as follows: Section \ref{sec:methods4search} provides an overview of the cWB-2G workflow and the key improvements introduced in preparation for O4, leading to the latest release of cWB-2G \cite{cWB-6.4.6}.
Section \ref{sec:dataset} describes the dataset used in this study and the configurations of the analyses, in particular the supervised training and testing procedures of the eXtreme Gradient Boosting (XGBoost) multivariate classifier.
Section \ref{sec:significance} presents a method for assessing the consistency of the empirical background model with a Poisson point process. This is crucial to check the statistics of false alarms provided by the machine-learning ranking.
In Section \ref{sec:FAcorrelations} we discuss the correlation between false alarms from the different searches investigated in this paper. Finally, in section \ref{sec:comparison_genericvstargeted} we compare and combine searches for generic GWTs exploiting two and three detectors and we  contrast an example of model--informed search for CBC coalescences with respect to model--agnostic searches.

\section{General overview of cWB-2G analysis scheme}
\label{sec:methods4search}
\begin{figure}
    \centering
    \includegraphics[width=0.8\textwidth]{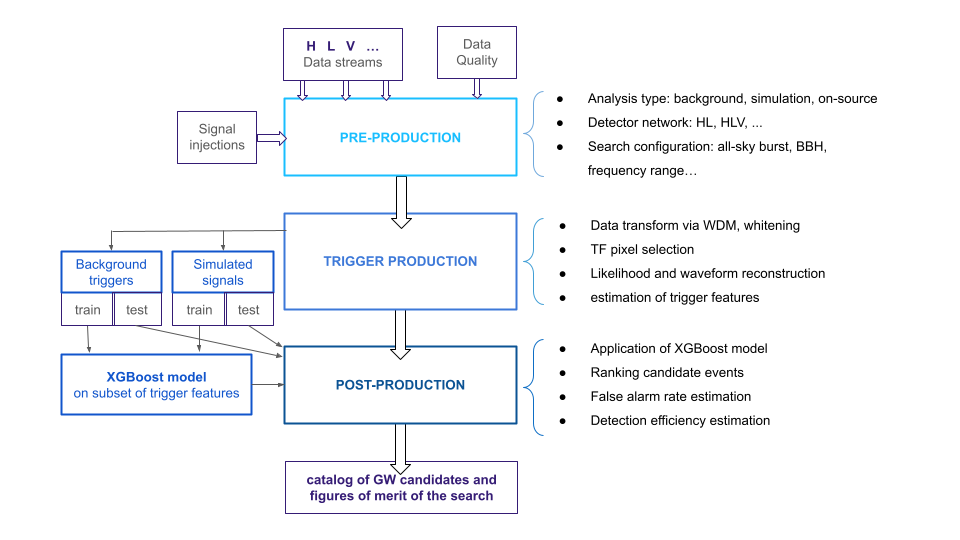}     
    \hfill
\caption{Workflow and data flow within cWB--2G.}
\label{fig:cwb_scheme}
\end{figure}

Coherent WaveBurst--2G \cite{cwb2016, cwbsftX, cwb_manual,cWB-6.4.6} is an analysis pipeline employed in searches of generic GWT signals, designed to operate with little or no prior knowledge of the waveform models.
Figure \ref{fig:cwb_scheme} shows a flowchart of cWB-2G which highlights the main logic blocks.
The analysis proceeds across four main stages: 
\begin{itemize}
    \item \textbf{pre-production}, where the configuration of the analysis is defined, e.g., which detector network is considered, type of search for GWTs (e.g. all-sky or follow-up, frequency range and time--frequency (TF) resolution levels), type of analysis (on-source analysis, estimation of accidental false alarms or simulation of software injections of GWTs), management of job submission;
    \item \textbf{production of candidate triggers} and their summary statistics. The first step is a data pre-conditioning procedure, which includes the mitigation of spectral lines by regression and the data whitening filter. This is followed by a preselection of trigger candidates in the time–frequency plane, primarily based on the search for coincident excess power.
    The identified triggers are then passed to the more computationally demanding stage of constrained likelihood maximization: in this step, the coherent response of two or more observatories is evaluated under the GWT signal hypothesis \cite{cwb2005, cwb2008}.
    The procedure further estimates the extrinsic and intrinsic properties of the candidate GWT, including its sky direction \cite{cwb2011} and signal waveform \cite{cWB2019_widerlook}, along with a set of features characterizing the selected trigger;
    \item \textbf{ranking of candidate triggers} by means of the XGBoost classifier, which has been previously trained on the trigger features provided by the background analysis and the simulation analysis, representing the accidental false alarms and the signal class, respectively. 
    \item \textbf{reporting results}, that delivers the catalog of candidate GWTs, their reconstructed features, and several figures of merit of the analysis.
\end{itemize}


In cWB-2G, the input strain data are whitened and analyzed in the Time--Frequency plane (TF) by the invertible orthonormal wavelet transform Wilson-Daubechies-Meyer (WDM) \cite{cwbWDM}. 
To preserve the sensitivity to diverse signal morphologies, cWB-2G utilizes concurrent WDM transforms at several different TF resolutions \cite{cwbWDM}. 
Once a trigger is identified as a potential GWT, cWB-2G reports several ad hoc statistics; we summarize here the most important ones.
The network correlation coefficient, $\text{c}_{\text{c}}$, evaluates the coherence of a trigger in the network of detectors, and is defined as \cite{cwb2016}
\begin{equation}
    c_c = \frac{E_c}{E_c + E_n} \, \text{,}
\label{eq:netcc0}
\end{equation}
where $E_n$ is the residual noise energy and $E_c$ is the coherent energy that quantifies the energy content of the GWT candidate.  Both $E_n$ and $E_c$ are estimated in whitened data as square of the corresponding signal-to-noise ratio (SNR) over the network of detectors.
The GWT detection statistics used through O3 LVK analyses is empirically defined as 
\cite{Mishra:2021tmu}
\begin{equation}
    \eta_0 = \sqrt{\frac{E_c}{1+\chi^2 \cdot \text{max}\left(1,\chi^2-1\right)}}
\label{eq:SNRnet}
\end{equation}
where $\chi^2 \equiv E_n/N_{df}$ is a proxy of the Chi-square statistics \cite{cwb_manual} and $N_{df}$ is the number of independent wavelet amplitudes used to characterize the trigger.

In the current LVK observing run O4, cWB-2G exploits the XGBoost machine learning (ML) algorithm \cite{XGBoost} to classify triggers according to their estimated authenticity, as explained in the following. The XGBoost algorithm is trained by learning false alarm and signal properties as described by a set of features, including $\eta_0$, $\text{c}_{\text{c}}$ and several others described in \ref{app:config}. 
The specific novelty of this cWB-2G version \cite{cWB-6.4.6} is that it ranks candidate triggers by relying solely on the output score provided by XGBoost, $ W_{\text{XGB}} \in [0,1]$, which is monotonically stretched to the new ranking statistics $\eta_0^{\prime}$ to mitigate numerical resolution limitations in the higher score range.  
This new automatic ranking eliminates the complex manual tuning that has been intrinsic in previous cWB-2G versions enhanced with ML \cite{Marek:2023}. In fact, the XGBoost scores were previously used as a penalty factor multiplying the traditional test statistic $\eta_0$. In addition, to mitigate short--duration false alarm glitches, known as \textit{blips} \cite{Cabero:2019orq}, that result was then fed in nonlinear penalization functions of a few morphological parameters of the candidate triggers. Such a complexity is not needed anymore with the current detection statistics $\eta_0^{\prime}$.

This methodological advancement is calling for more extended tests and thorough validation procedures to exclude systematics, as they might emerge from e.g. ML overfitting or correlations in the data set representing false alarms. 
Further details of the tuning of the XGBoost architecture, including the choices on the benchmarks of performances, are discussed in \ref{subsec:tuningXGBoost}, while the training procedures used for different search types are discussed in Section \ref{sec:dataset} with more details in \ref{subsec:trainingXGBoost}. Section \ref{sec:significance} and \ref{app:BKG} discuss new validation methods implemented concerning the accuracy of the assessment of the statistical significance of candidate GWTs.



\section{Data set and configuration of the analyses}
\label{sec:dataset}
In this work, we compare different cWB-2G analyses of publicly available data from the second half of the LVK third observing run (O3b \cite{OpenData_2023}) of advanced LIGO \cite{LIGO-O3} and advanced Virgo detectors \cite{VirgoDetCharO3}, more specifically data from 2020, January 6 (GPS 1262304000 s) to February 14 (GPS 1265760000). Within this time range, there is a good coverage of coincident observation by the three detectors, corresponding to about 14.8 days of science quality data. We restrict analyses to these data, considering separately the two detector network made by LIGO-Hanford (H) and LIGO-Livingston (L) and the three detector network including Virgo (V). The duration of this data set is well-suited to the methodological goals of this work, and the associated computational load is not excessive.

Three different all-sky analyses are then developed by training different models of XGBoost post-processing:
\begin{itemize}
    \item search of short duration GWTs on HL data (HL-burst);
    \item search of short duration GWTs on HLV data (HLV-burst);
    \item search of GWTs from binary black hole coalescences on HL data (HL-BBH);
\end{itemize}
All these cWB-2G analyses are performed in the frequency range $[16-2048]$ Hz, which is the same choice adopted in off-line searches for GWTs on O4 data, motivated by general considerations on the noise power spectral densities of the detectors and on the predicted spectral content of interesting sources such as CCSN, which may extend up to 2 kHz. 
The analyses employ the WDM transform \cite{cwbWDM} with seven time--frequency resolutions with $\Delta t \times \Delta f$  in the range from $[256 \si{\hertz} \times 1.95 \si{\milli\second}] $ to  $[4 \si{\hertz} \times 125 \si{\milli\second}] $. 
The sky directions are analyzed using segmentation Healpix level 5, corresponding to a resolution of solid angle $\sim 3.4 \text{deg}^2$, which is fully sufficient for detection purposes.

The set of trigger features which are fed into XGBoost is search-dependent and are all estimated in whitened data (more details in \ref{app:config}). In the case of short burst searches, the input features include generic characteristics of triggers such as $\eta_0$, $\text{c}_{\text{c}}$ as well as morphological information inspired to the characteristics of the dominant glitch populations, such as the similarity of a trigger to a single pulse and the effective number of signal cycles (see Tab.\ref{tab:xgboost_features}). The latter features are especially useful to mitigate the effects of the so-called blip glitches \cite{Cabero:2019orq, VirgoDetCharO3}.
Information like estimated central frequency, duration and frequency bandwidth are not passed as features, with the aim to preserve the widest possible prior on the signal parameter space. The list of features is the same for the HL-burst and HLV-burst, but each feature is the result of the coherent reconstruction of the signal candidate in the different network of detectors.

In the search for BBH signals, the list of features of the short burst search is extended to include morphological characteristics that are model-specific to the target signal population, including estimates of chirp mass, similarity to chirping signal morphology, central frequency, frequency bandwidth, and duration (see Tab. \ref{tab:xgboost_features_BBH}). 
The HL-BBH trigger list is identical to that of HL-burst; its specialization arises solely from the dedicated ranking process conducted using XGBoost.

\begin{figure}
    \centering
    \includegraphics[width=0.7\textwidth]{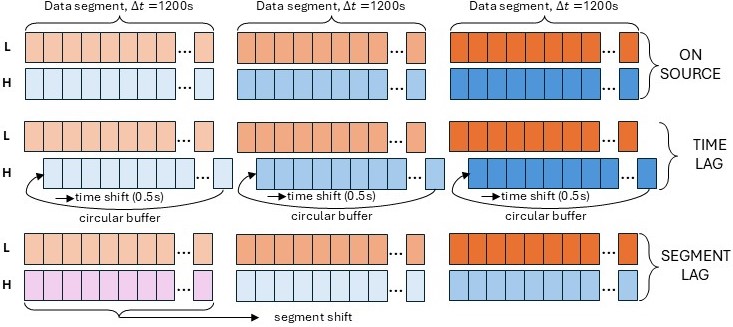}     
    \hfill
\caption{Implementation of time-slides in cWB-2G for two detectors: schematic view of three adjacent data segments. Top: actual data streams. Middle: time-lag treating each data segment as a circular buffer. Bottom: time shift by one data segment.}
\label{fig:cwb_time_shift}
\end{figure}

\subsection{Background triggers for training and testing the XGBoost classifier.}
\label{subsec:bkg}
The background triggers used to train and test the XGBoost classifier are provided by running cWB-2G over time-slides. Each time-slide is implemented by shifting in time the detectors' data streams with a set of {\it unique} time shift values, meaning that they are selected avoiding any repetition of the same time difference between detector pairs. The live-times of different time-slides are typically varying by a few \%, because the overlap of science quality data from different detectors changes with the applied time shifts.

For computational efficiency, cWB-2G analyzes the detectors' data streams in non-overlapping segments of 1200 s duration. Time-slides are implemented by combining two procedures: local time shifts within each data segment and segment--lags that shift data by multiples of the segments' duration, see Fig. \ref{fig:cwb_time_shift}.
Local time shifts are implemented as multiples of $0.5 \si{\second}$ and treat each data segment separately as a circular buffer. The $0.5\, \si{\second}$ step is enough to cancel the coherence of GWTs and noise outliers in the network. These local time-slides do not change the average time shift between detector pairs and provide up to 2400 instances of the noise.\footnote{Up to 2399 instances of noise triggers in case the segment-lag is zero.}
Differently, the segment-lags procedure changes which data segments are overlapped and contributes a net average time shift between detector pairs which is a multiple of $1200\, \si{\second}$. 


In this work we use up to 29 segment-lags, corresponding to average time--shifts from $-16800 \si{\second}$ to $+16800 \si{\second}$, for a total of up to 69599 time-lags. This is suitable to investigate false alarm probabilities at $\sim$ a few tens per million and to check for possible systematics related to day-time activities at the detectors' sites.\footnote{In previous cWB analyses of the LIGO observatory, the number of implemented time-lags has been beyond 1 million, see e.g. \cite{GW150914minimalassumptions}.} 
By randomly choosing half of these time-lags, we obtain the set of background triggers used as noise model for training XGBoost, while the complementary set is used for testing, i.e. to measure the statistical significance of candidate events. 

\begin{figure}
    \centering
    \includegraphics[width=0.7\textwidth]{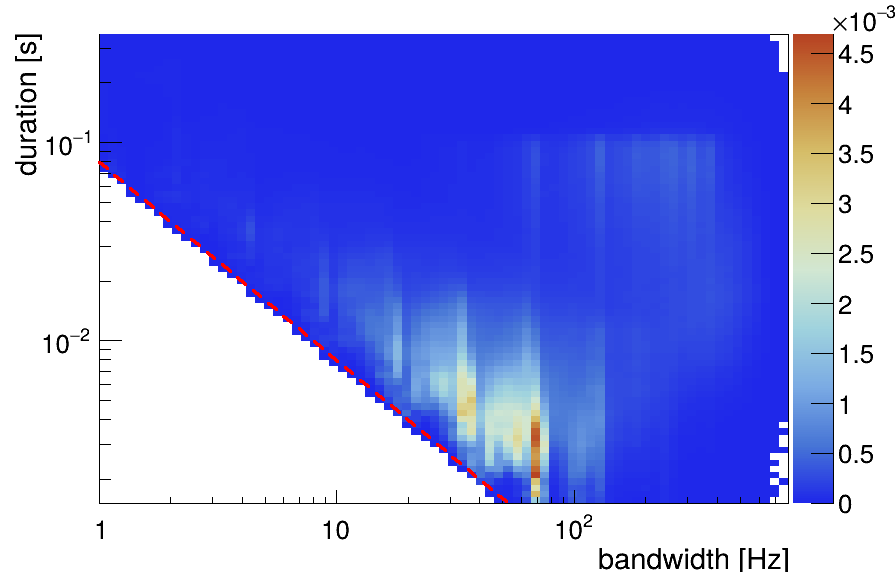}     
    \hfill
\caption{Distribution of duration and bandwidth of background triggers reconstructed by cWB-2G on time--lags of HL whitened data. 
Color scale shows fraction of counts per bin, that are uniform in log scale. The red dashed line shows the Heisenberg--Gabor limit for the time-frequency representation of triggers. }
\label{fig:bkg_cwb}
\end{figure}

Fig.\ref{fig:bkg_cwb} shows the distribution of the background triggers reconstructed by cWB-2G on time-lags of HL whitened data in the duration vs frequency bandwidth plane. 
The minimum amplitude Signal--to--Noise Ratio, SNR, which enables the reconstruction of a trigger is $\sim 5$.
Duration and bandwidth are estimated as standard deviations of the $\text{SNR}^2$ of each trigger in time and frequency.
This set of a few millions background triggers is then fed into the XGBoost classifier, half for training and half for testing, to implement the HL--burst and HL--BBH searches. 
The distribution is dominated by short duration background triggers, 
close to the Heisenberg-Gabor bound on the time-frequency representation of triggers, i.e. product of standard deviations of time and frequency coordinates $\sigma_t \sigma_f \geq \frac{1}{4\pi}$. 
The prevalence of such background triggers is not surprising. In fact, the detection efficiency is typically higher for triggers with a compact time-frequency representation while the rejection of accidental coincidences is typically better for triggers showing more complex morphologies, thanks to the coherent analysis over the network of detectors. In addition, we know that short-duration noise outliers are very common in each detector.
A very similar distribution of background triggers is obtained by the analysis of HLV time-lags, and it is used for tuning the HLV--burst search. 
The distributions of the $\sim 10^5$ most significant background triggers according to the three different searches, as established by the respective XGBoost ranks, are discussed in sec.\ref{sec:significance}.

\begin{figure}
\begin{subfigure}[b]{0.495\textwidth}
    \centering
        \includegraphics[width=\textwidth]{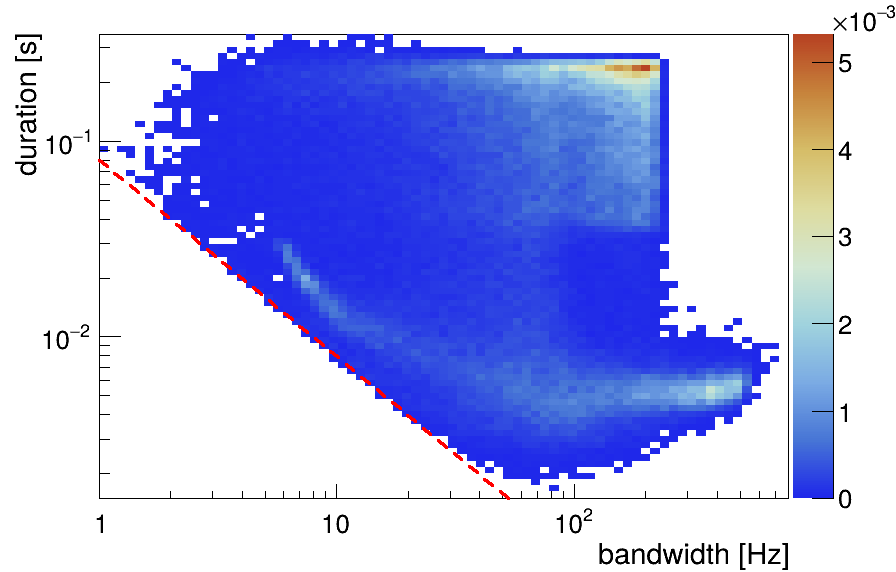}
        \caption{GWT set for training, WNBs}
        \label{fig:simTrain}
    \end{subfigure}
    \hfill
    \begin{subfigure}[b]{0.495\textwidth}
        \centering
        \includegraphics[width=\textwidth]{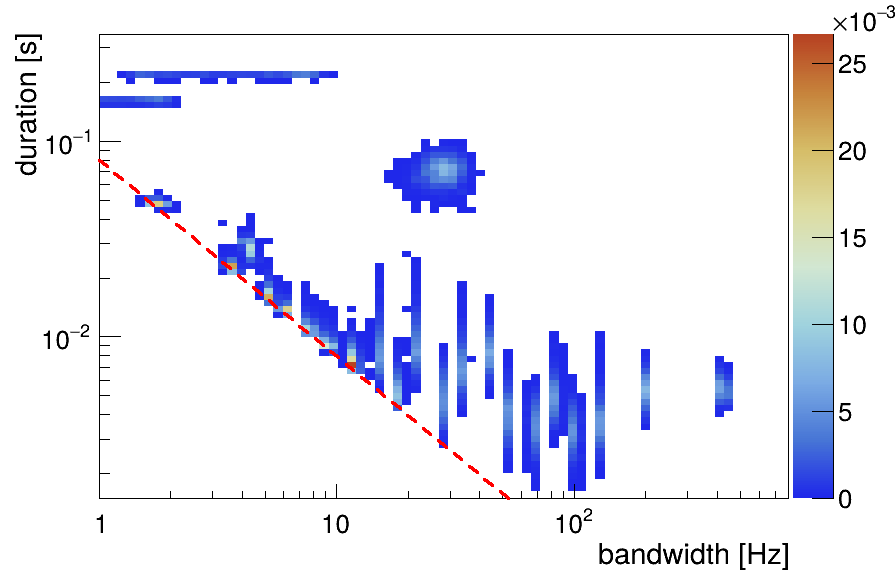}
        \caption{GWT set for testing, mixed morphologies}
        \label{fig:simTest}
\end{subfigure}
\caption{Distribution of reconstructed duration and bandwidth for the injected signals, zero--noise and whitened, used as training and testing sets in HL-burst and HLV-burst.
The red dashed line shows the Heisenberg--Gabor limit. Color scale indicates fraction of counts per bin.
Panel (a): training set of WNBs. 
Panel (b): testing set of ad--hoc waveform morphologies - sine-Gaussians near the Heisenberg--Gabor limit, shorter--duration Gaussian pulses, WNBs with larger time-frequency volumes.}
\end{figure}

\subsection{Gravitational wave signals for training and testing the XGBoost classifier.}
\label{subsec:trainingset}

The simulated GWTs used for training and testing procedures are injected in GW strain data and analyzed by cWB-2G. The populations of simulated signals are chosen according to the goals of each search. All-sky searches for short duration GWTs of generic morphology lack a well-defined signal model. Therefore, the optimization of signal sets for training and testing HL-burst and HLV-burst is ill-defined, and their selection is based on qualitative criteria. It is particularly important that the training and testing populations differ sufficiently, with the former lacking distinctive morphological features, in order to assess the algorithm’s ability to generalize to previously unseen signals.  
Instead, model-informed searches, such as HL-BBH, can exploit the same GWT population model for training and testing.

The qualitative criteria adopted for the selection of simulated GWTs to train the HL-burst and HLV-burst searches include:

\begin{itemize}
    \item lower amplitude GWTs are likely more abundant than higher amplitude ones;
    \item all sky directions and arrival times during the observation should be uniformly distributed;
    \item preserve sensitivity to diverse GWTs, without  assuming patterns in the signals' waveforms.
\end{itemize}

HL-burst and HLV-burst are trained using white noise burst (WNB) injections, which are GWTs with random polarization amplitudes as described in \cite{cwb2023}. WNBs provide a challenging test case due to their noise--like properties. Their central frequency, bandwidth and duration parameters are randomly sampled across the searched ranges, in particular the duration and bandwidth span approximately from 2 ms to 0.3 s and from 2 Hz to 0.5 kHz respectively. 
For the signal model, duration and bandwidth are defined as standard deviations in time and frequency of the $\text{SNR}^{~2}$ of the injected waveform after whitening, i.e. in zero--noise condition. The resulting distribution of reconstructed bandwidth and duration of these WNBs is shown in Fig. \ref{fig:simTrain}.
The chosen population emphasizes training signals with larger time--frequency spread, since their lower SNR density makes them harder to detect. We also enhance the population of signals near the Heisenberg-Gabor limit to match time-frequency characteristics of the dominant background triggers (see Fig.\ref{fig:bkg_cwb}). The amplitudes of WNBs are simulated from a distribution $\propto  \text{SNR}^{-2}$, down to a minimum SNR in the network of detectors that corresponds to a negligible detection efficiency. 
This amplitude distribution enhances the training to signals along directions of weaker sensitivity of the network of detectors, consistent with the aim to make the sky coverage as isotropic as possible. This choice achieves very good results for diverse populations of sources, e.g. either uniformly distributed in space or else uniformly distributed in distance.

The signal test set comprises diverse GWTs morphologies traditionally used by LVK collaborations for burst searches \cite{allsky1,allsky2,allsky3}. These include:

\begin{itemize}
    \item Sine--Gaussian waveforms with elliptical (SGE) and linear (SG) polarization, at various central frequencies and Q--factors (3, 9, and 100), providing signals with compact time--frequency representations;
    \item WNBs at a few central frequencies, with random polarization amplitudes and larger time--frequency volumes;
    \item Linearly polarized Gaussian pulses (GA), short duration (from 0.1 to 4 ms) and  with larger bandwidth, similar to the effective bandwidth set by the noise power spectral density of detectors. 
\end{itemize}

\begin{figure}
    \begin{subfigure}[b]{0.495\textwidth}
    \centering
    \includegraphics[width=\textwidth]{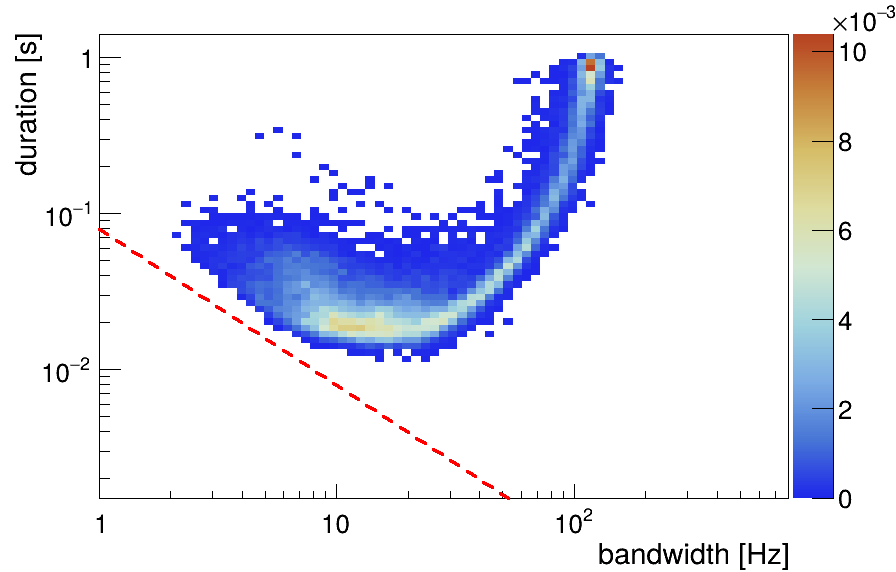}
    \caption{injected CBC signals}
    \label{fig:simCBCTest_mod}
\end{subfigure}
   \begin{subfigure}[b]{0.495\textwidth}
    \centering
    \includegraphics[width=\textwidth]{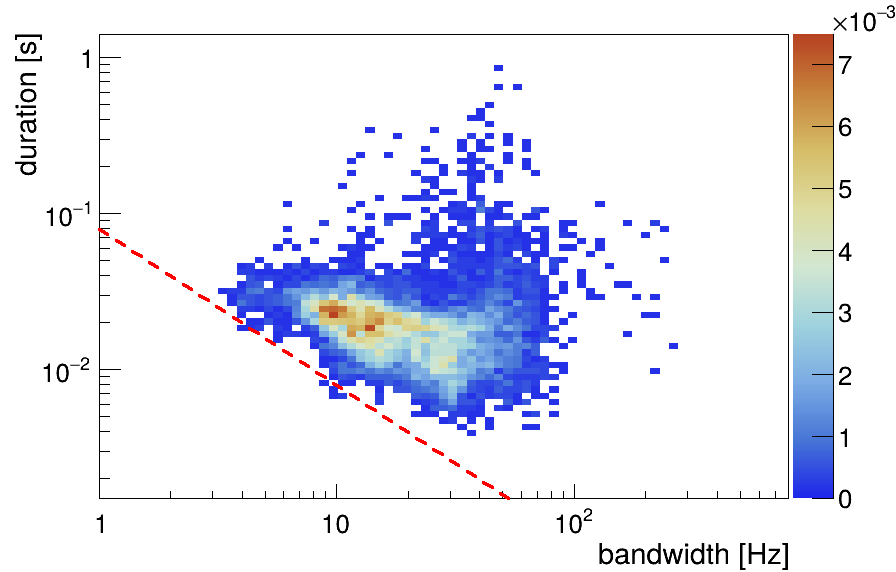}
    \caption{recovered CBC signals}
    \label{fig:simCBCTest_rec}
\end{subfigure}
\caption{Distribution of CBC signals used for training
and testing the HL--BBH search, in the duration-bandwidth plane. Panel (a) shows the simulated signals, zero--noise and whitened, 
with BNS and NS-BH signals appearing at longer durations, $\sim$ 1s. Panel (b) shows the whitened signals found by cWB 2G, which are then passed to the XGBoost classifier. 
Color scale shows the fraction of counts per bin. The red dashed line indicates the Heisenberg--Gabor limit.}
\label{fig:simCBCTest}
 \end{figure}

The distribution of reconstructed duration and bandwidth of these test signals is shown in Fig. \ref{fig:simTest}: clearly, they are just sparsely sampling the space of detectable short duration GWTs. More signal classes are usually added to extend the testing and the interpretation of the search; in fact, we also evaluate the sensitivity of HL-burst and HLV-burst using simulated CCSN signals and the model of CBC sources described below and shown in Fig.\ref{fig:simCBCTest_mod}. Other aspects related to the distribution of simulated sources are discussed in Sec.\ref{sec:comparison}. 

\vskip\baselineskip

For the model--informed search HL-BBH, we use  distinct sets of $\simeq 10^5$ injections each for training and testing, sampled from the Power Law + Peak mass distribution derived from LVK observing runs O1-O3 \cite{RateAndPopO3}.
Fig. \ref{fig:simCBCTest} shows the distribution of signal duration and bandwidth of whitened signals, with panel (a) showing the injected ones, zero-noise, and panel (b) the found ones, as recovered by cWB-2G from the noisy data streams. The comparison between panels and further investigations reveal that HL-BBH has a lower sensitivity to BNS and NS-BH coalescences, which have longer duration and larger bandwidth, with standard deviations $\sim$ 1 s and $\sim$ 100 Hz in Fig. \ref{fig:simCBCTest}. This limitation could be mitigated with a CBC-specific configuration of cWB-2G, in particular by tuning the trigger preselection and coherent analysis algorithms, but here we chose to focus on the effects of the tuning of the XGBoost post-processing and we kept the same cWB-2G configuration of the HL-burst. 
Overall, the major cause of missed CBC detections is by far related to the more distant sources in the simulated population. Subsec. \ref{subsec:BBH_burst_SIM} discusses the resulting sensitivity to BBH and IMBH mergers.  

\section{Properties of false alarms and statistical significance of GWT candidates}
\label{sec:significance}

\vskip\baselineskip
\begin{figure}
\begin{subfigure}[b]{0.328\textwidth}
        \centering
        \includegraphics[width=\textwidth]{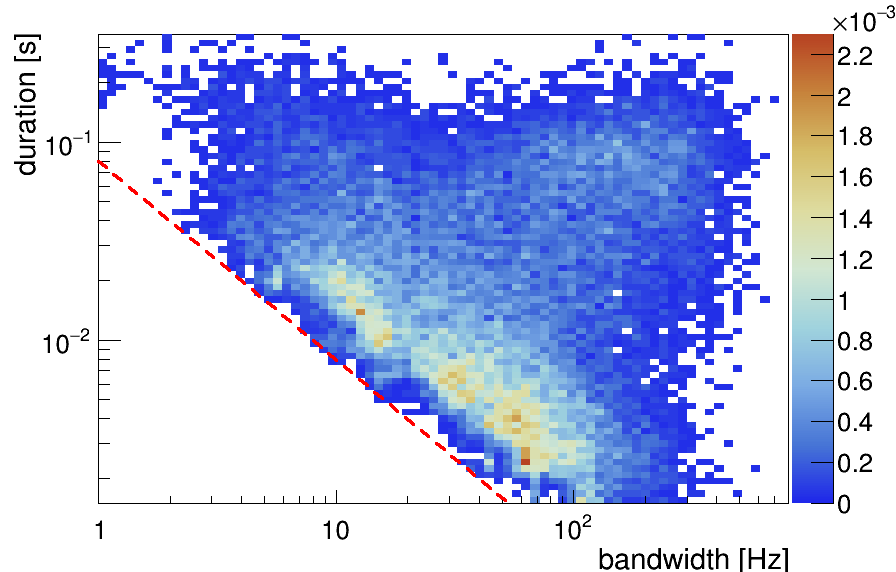}
        \caption{HL-burst}
        \label{fig:bkg_xgb}
\end{subfigure} 
\hfill
\begin{subfigure}[b]{0.328\textwidth}
        \centering
        \includegraphics[width=\textwidth]{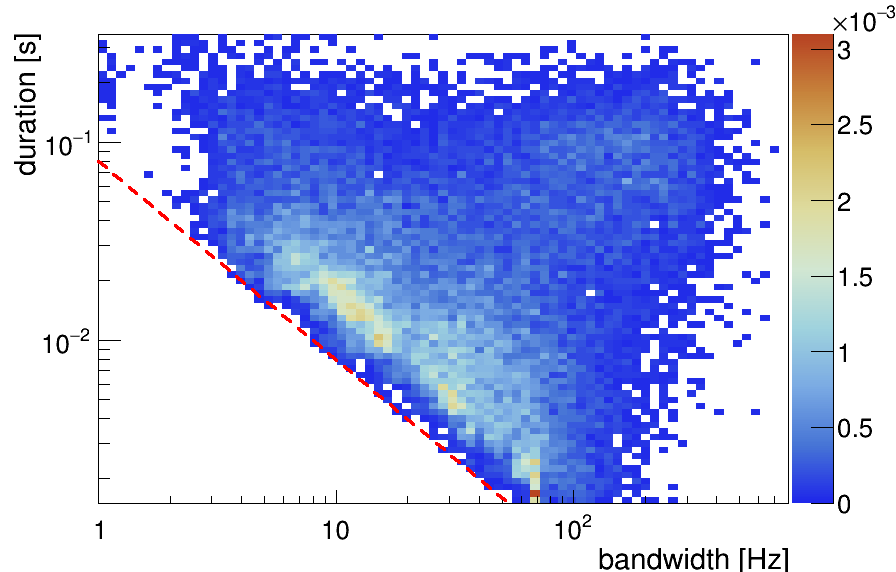}
        \caption{HLV-burst}
        \label{fig:bkg_xgb_LHV}
\end{subfigure} 
\hfill
\begin{subfigure}[b]{0.328\textwidth}
     \centering
      \includegraphics[width=\textwidth]{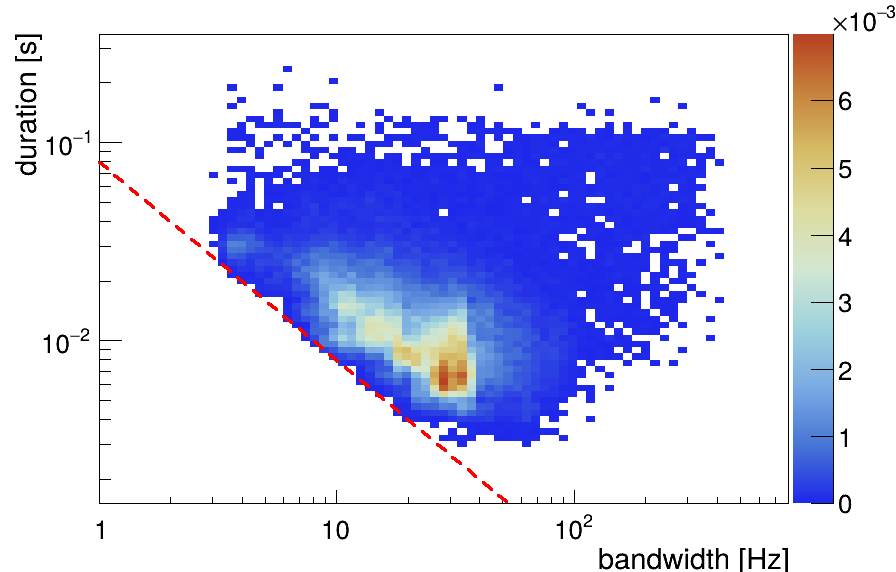}
        \caption{HL-BBH}
        \label{fig:bkg_bbh}
\end{subfigure}
\caption{Distributions of duration and bandwidth for the background triggers of the three searches passing an IFAR threshold of 1 week: (a) HL-burst, (b) HLV-burst, (c) HL-BBH. Color scales report the false alarm fraction per bin. The total count of false alarms ($\sim 10^5$) and the binning are the same in the three panels. We recall that the different false alarm sets of HL-burst and HL-BBH are resulting from the implementation of different XGBoost ranking procedures on the same set of background triggers found by cWB-2G from HL time-slides. The red dashed line indicates the Heisenberg--Gabor limit. 
}
\label{fig:bkg_search}
\end{figure}

The background triggers found by the three searches (HL-burst, HLV-burst and HL-BBH) show a variety of morphological features. Fig. \ref{fig:bkg_search} compares their distributions as a function of duration and bandwidth, i.e. standard deviations of $\text{SNR}^{~2}$ in time and frequency, for all accidental triggers passing the Inverse False Alarm Rate (IFAR) threshold of 1 week.
In all three searches, accidental triggers with a compact time-frequency morphology are still dominating the distributions, even though the ranking provided by XGBoost is mitigating the impact of the shorter duration triggers compared to the cWB-2G background shown in Fig. \ref{fig:bkg_cwb}.
HL-burst and HLV--burst show similar distributions of accidental triggers, see e.g. Fig. \ref{fig:bkg_xgb} and \ref{fig:bkg_xgb_LHV}. In addition to the population which is peaked close to the Heisenberg-Gabor limit, a secondary component is visible with standard deviations in duration and bandwidth $\sim 0.1$ s and a few hundred Hz respectively. We recall that, in HL-burst and HLV-burst searches, the duration and bandwidth features are not passed to the XGBoost classifier to preserve their agnostic character. 
Background triggers from HL-BBH are more compact in the duration-bandwidth plane, see Fig. \ref{fig:bkg_bbh}. This is expected, given the properties of targeted signals shown in Fig. \ref{fig:simCBCTest_mod} and the exploitation by XGBoost of more features of reconstructed triggers, including here also the duration and bandwidth. 
However, the resulting mitigation of the dominating population of background triggers close to the Heisenberg-Gabor limit is still partial, mainly because of the residual overlap between distributions of estimated features of accidental triggers and of detected GWT injections.

\label{subsec:FAconsistency}  
The statistical significance of GWT candidates is estimated by comparison with the empirical distribution of false alarms from time-slides. This procedure can be affected by systematic errors on the point estimate and its uncertainty, that may be caused by e.g. correlations in the set of background triggers, non--stationary noise, residual effects due to the presence of genuine GWTs in the data, and finite sample size \cite{Was2010}. Given the novelty of the detection statistic adopted here, it is  important to provide a thorough check of the accuracy of statistical significance estimates as well as to test if the fluctuations are consistent with a homogeneous Poisson point model.
The expected counts of background triggers is numerically simulated by assuming a Poisson distribution of the counts per each time-lag $\mathcal{P}(k; \mu )= e^{-\mu} \mu^k / k!$, see \ref{app:BKG}. Here, the expectation value $\mu$ is set equal to $\text{FAR}(\eta_0^{\prime}) \ T $, where $T$ is the live-time of each time-lag and $\text{FAR}(\eta_0^{\prime})$ is the measured value of the false alarm rate at threshold $\eta_0^{\prime}$ on the detection statistic.

\begin{figure}
\centering
\includegraphics[width=0.8\textwidth]{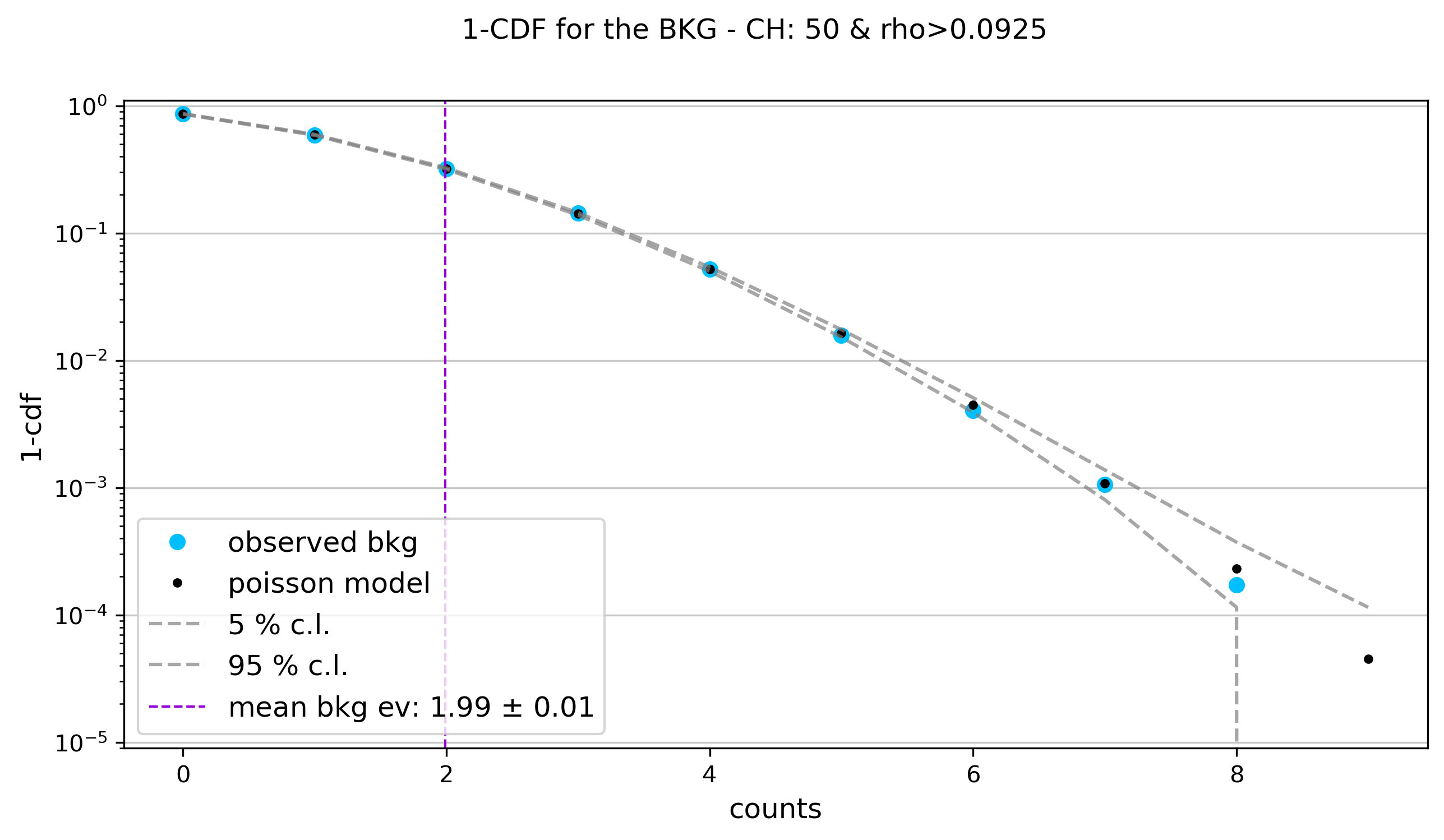}
\caption{Cumulative distribution of false alarms counts, as empirically estimated from time slides, vs expectations from the Poisson model, in a time--lag of HL--burst at an IFAR thresholds of $\simeq$ 1 week. The colored markers report the frequencies of occurrence of a false alarm count larger than the integer $k$ in abscissa; the false alarm count zeroes at $k=$9. The black markers and dashed curves represent the Poisson expectations for the probability (mean with 5\% and 95\% percentiles). The vertical dashed line indicates the mean counts per time-lag at this threshold.
There is full agreement over all the investigated range of integer counts. Similar agreement is achieved in all three searches at IFAR thresholds from $\sim$ 1 day to $\sim$ 1 century.}
\label{fig:PoissonCDF}
\end{figure}

The distribution of false alarm counts from time slides is in good agreement with the Poisson model for the three searches over a wide range of IFAR thresholds, from $\sim$ 1 day to $\sim$ 1 century. 
Fig.\ref{fig:PoissonCDF} gives an example of the good agreement for an IFAR threshold of 1 week in HL-burst: the frequencies of occurrence of counts of false alarms above threshold are 
consistent with the Poisson model tuned to the empirical mean rate of events at that threshold. This agreement is confirmed for all searches in the entire range of tested IFAR thresholds, down to cumulative probabilities as low as a few $10^{-5}$, limited by the total number of time slides performed in this work. 
Tab.\ref{tab:PoissonCDF} reports a summary of the agreement in probability of non-zero counts over the wide range of tested IFAR thresholds. From these checks, we can conclude that there is no evidence for systematic biases of our point estimates of false alarm probabilities, nor for systematic deviations from the Poisson model. 

\begin{table}
\footnotesize
\lineup
\begin{tabular}{c l c c c}
\br

\multicolumn{5}{c}{HL--burst search} \\
\mr                            
$\text{threshold}$ &
$\text{FAR}_{\text{emp}} \pm \sigma^{\text{FAR}}_{\text{emp}}$ &
$\mathcal{P}_{\text{emp}}(\text{counts} > 0)$ &
$\mathcal{P}_{\text{th}}(\text{counts} > 0)$ &
$\sigma_{\text{th}}$ \cr
\mr
$\text{A}$ & $(1.001 \pm 0.001) /\text{day}$      & 1.0     & 1.0 & $5\cdot10^{-6}$ \cr
$\text{B}$ & $(1.000 \pm 0.004) /\text{week}$     & 0.864   & 0.863    & 0.002           \cr
$\text{C}$ & $(1.01  \pm 0.02)  /\text{6 months}$ & 0.074   & 0.074    & 0.001           \cr
$\text{D}$ & $(0.98  \pm 0.03)  /\text{year}$     & 0.037   & 0.037    & 0.001           \cr
$\text{E}$ & $(0.99  \pm 0.09)  /\text{10 year}$  & 0.0038  & 0.0038   & 0.0003          \cr
$\text{F}$ & $(0.9   \pm  0.3)  /\text{100 year}$ & 0.00034 & 0.00035  & 0.0001          \cr
\br

\multicolumn{5}{c}{HLV--burst search} \\
\mr                            
$\text{threshold}$ &
$\text{FAR}_{\text{emp}} \pm \sigma^{\text{FAR}}_{\text{th}}$ &
$\mathcal{P}_{\text{emp}}(\text{counts} > 0)$ &
$\mathcal{P}_{\text{th}}(\text{counts} > 0)$ &
$\sigma_{\text{th}}$ \cr
\mr
$\text{A}$ & $( 1.000 \pm 0.002 ) /\text{day}$      & 1.0    & 0.999999 & $7\cdot10^{-6}$  \cr
$\text{B}$ & $( 0.999 \pm 0.006 ) /\text{week}$     & 0.871  & 0.868    & 0.003            \cr
$\text{C}$ & $( 1.00  \pm 0.03 )  /\text{6 month}$  & 0.075  & 0.075    & 0.002            \cr
$\text{D}$ & $( 1.00  \pm 0.04 )  /\text{year}$     & 0.038  & 0.038    & 0.002            \cr
$\text{E}$ & $( 1.0   \pm 0.1 )   /\text{10 year}$  & 0.0039 & 0.0039   & 0.0005           \cr
$\text{F}$ & $( 0.9   \pm 0.4 )   /\text{100 year}$ & 0.0003 & 0.0003   & 0.0002           \cr
\br

\multicolumn{5}{c}{HL--BBH search} \\
\mr                            
$\text{threshold}$ &
$\text{FAR}_{\text{emp}} \pm \sigma^{\text{FAR}}_{\text{emp}}$ &
$\mathcal{P}_{\text{emp}}(\text{counts} > 0)$ &
$\mathcal{P}_{\text{th}}(\text{counts} > 0)$ &
$\sigma_{\text{th}}$ \cr
\mr
$\text{B}$ & $(1.000 \pm 0.004) /\text{week}$     & 0.865  & 0.863  & 0.002  \cr
$\text{C}$ & $(1.00  \pm 0.02)  /\text{6 moths}$  & 0.074  & 0.074  & 0.001  \cr
$\text{D}$ & $(0.99  \pm 0.03)  /\text{year}$     & 0.038  & 0.037  & 0.001  \cr
$\text{E}$ & $(1.00  \pm 0.09)  /\text{10 year}$  & 0.0038 & 0.0038 & 0.0003 \cr
$\text{F}$ & $(0.9   \pm 0.3)   /\text{100 year}$ & 0.0003 & 0.0003 & 0.0001 \cr
\br

\end{tabular}
\caption{Probability for a non-zero count of false alarms in one time--lag at selected FAR thresholds from $\simeq$ 1/d, to 1/100 yr, labeled from A to F. From top to bottom: results of the HL--burst, HLV--burst and HL--BBH The empirical estimates from time--slides (3$^{rd}$ column) are consistent with expectations from the Poisson model (4$^{th}$ column). The last column reports the standard deviation from the Poisson model.
}
\label{tab:PoissonCDF}
\end{table}

\begin{figure}
\centering
\includegraphics[width=0.8\textwidth]{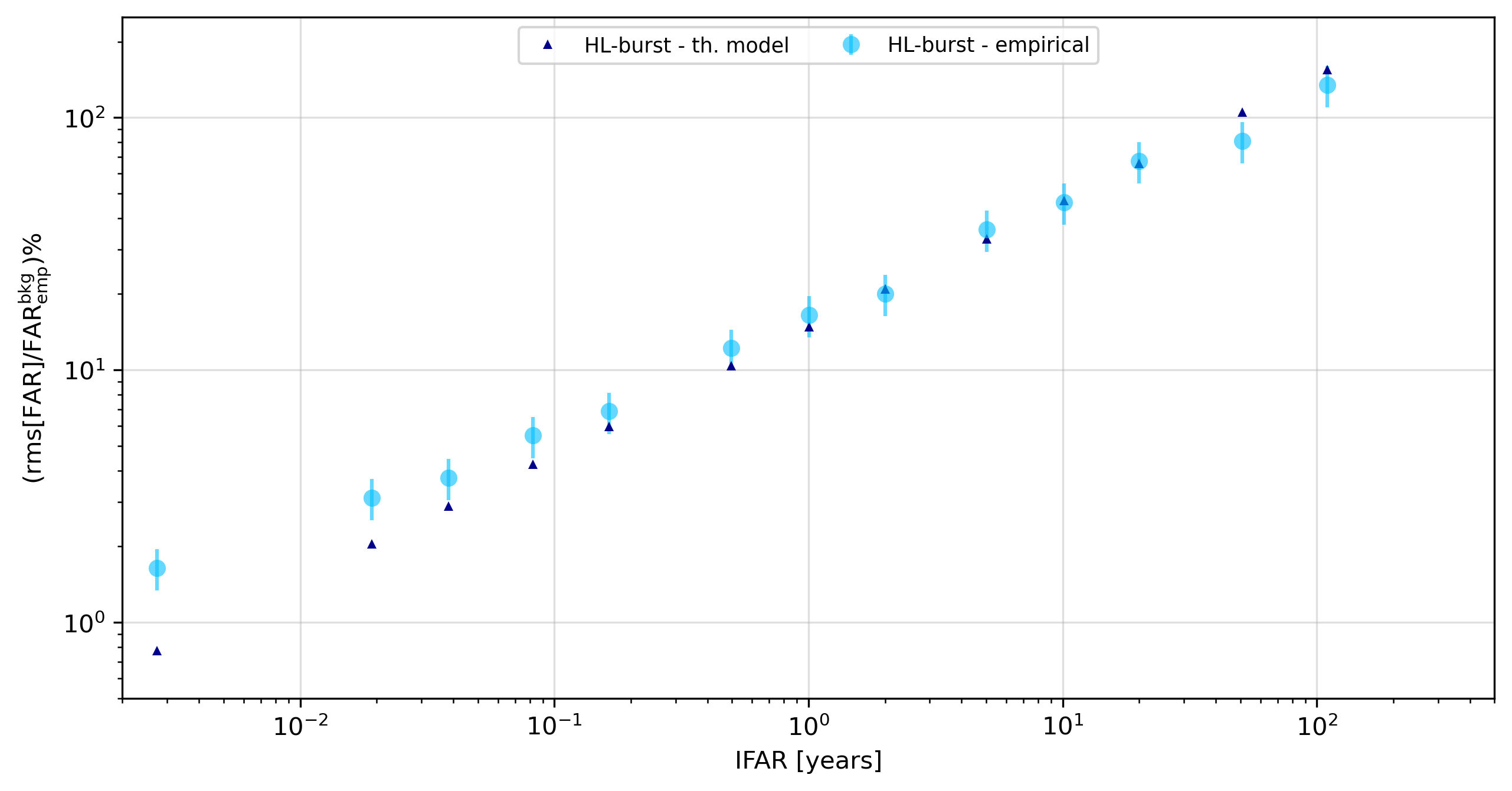}
\caption{Root-mean-square (RMS) of relative fluctuations of empirical FAR estimates from 29 separate segment-shifts of LIGO Hanford detector as a function of threshold values for the HL--burst search (blue markers with $\pm 1 \sigma$ error bars). 
The predicted relative fluctuations from the numerical Poisson model at the same IFAR values are plotted as black triangles.}
\label{fig:FAR_rms}
\end{figure}

In non-stationary noise, as e.g. related to human activities, one may expect a correlation of the rate of false alarms with the time shift between detectors. We analyzed separately the FAR estimates per each segment--shift at fixed values of the ranking statistics $\eta_0^{\prime}$, covering time shifts of LIGO Hanford data in the range [-4h40min, +4h40min] in steps of 20 min, see subsec. \ref{subsec:bkg} and \ref{app:BKG}.\footnote{Here, a time shift of -2h synchronizes data streams to the same local time at Hanford and Livingston sites.}  
The resulting point estimates of FAR do not show a correlation with the time--shift applied between the LIGO detectors.\footnote{We remark that the time shift applied to the Virgo detector plays a less relevant role on the statistic of false alarm for HLV analyses, given that for most sky directions and GW polarizations the network SNR is dominated by HL.}  
Moreover, we also tested the fluctuations of FAR point estimates from different segment-shifts and found that they are explained by the Poisson model at IFAR thresholds of 2 months or larger, see Fig.\ref{fig:FAR_rms}. 
The fluctuations exceed the poissonian statistical uncertainties only at lower IFAR thresholds and are anyway limited to $<$ 5\%; therefore they can be neglected in the assessment of the statistical significance of GWT candidates. 

Summarizing, these investigations brought no evidence for the presence of relevant systematic effects in our estimates of the statistical significance of GTW candidates in our searches on O3 data. 

\section{Correlations of background triggers in different searches}
\label{sec:FAcorrelations}

A key question in combining results from different searches is to take into account the joint significance of candidate GWTs. This section focuses on the correlation between background triggers across different searches, more specifically on the probability that a background trigger is in common to more searches. 
We compare false alarm results from pairs of searches using the same subset of time slides, more specifically, analyzing only common time--shifted live--times. In this condition, the probability that a background trigger in search A also appears in search B, $\mathcal{P}(B|A)$, is estimated by the count of common false alarms divided by the count of false alarms from search A. 
In case searches A and B are analyzing the same observing time and are set at the same IFAR threshold, 
$\mathcal{P}(B|A)$ is equal to $\mathcal{P}(A|B)$ by construction, since the two sets of background triggers will have the same expected counts, i.e. the same cardinality, and $\mathcal{P}(A)=\mathcal{P}(B)$.
Therefore, the effective IFAR for the combined search made by the logical OR of two concurrent searches run at the same $\text{IFAR}^{\prime}$ threshold will be $\text{IFAR}=\text{IFAR}^{\prime} / (2 - \mathcal{P}(A | B))$, where the denominator is the effective trials factor. 


\begin{table}[]
\centering
\footnotesize
\renewcommand{\arraystretch}{1.2}
\begin{tabular}{c c c c | c c c}
\hline
\multicolumn{7}{c}{False alarm counts}  \\
& \multicolumn{3}{c}{HL-burst vs HLV-burst} & \multicolumn{3}{c}{HL-burst vs HL-BBH} \\
\br
Treshold & common & exclusive HL & \multicolumn{1}{c|}{exclusive HLV} & common & exclusive burst & exclusive BBH \\
\br
A & 3092 & 14533 & 14412 & -   & -    & -    \cr       
B & 470  & 2042  & 2034  & 126 & 2386 & 2479 \cr  
C & 59   & 229   & 234   & 9   & 279  & 283  \cr  
D & 7    & 41    & 46    & 0   & 48   & 50   \cr  
E & 2    & 4     & 2     & 0   & 4    & 5    \cr
\hline
\end{tabular}
\caption{Counts of background triggers at different search thresholds (from A to E, see Tab. \ref{tab:PoissonCDF}) on the same time-lags between H and L detectors: common and exclusive false alarms in HL-burst vs. HLV-burst (columns 2-4) and in HL-burst vs.  HL-BBH (columns 5-7).
}
\label{tab:FAcomparison}
\end{table}

\begin{figure}
\centering
\includegraphics[width=0.8\textwidth]{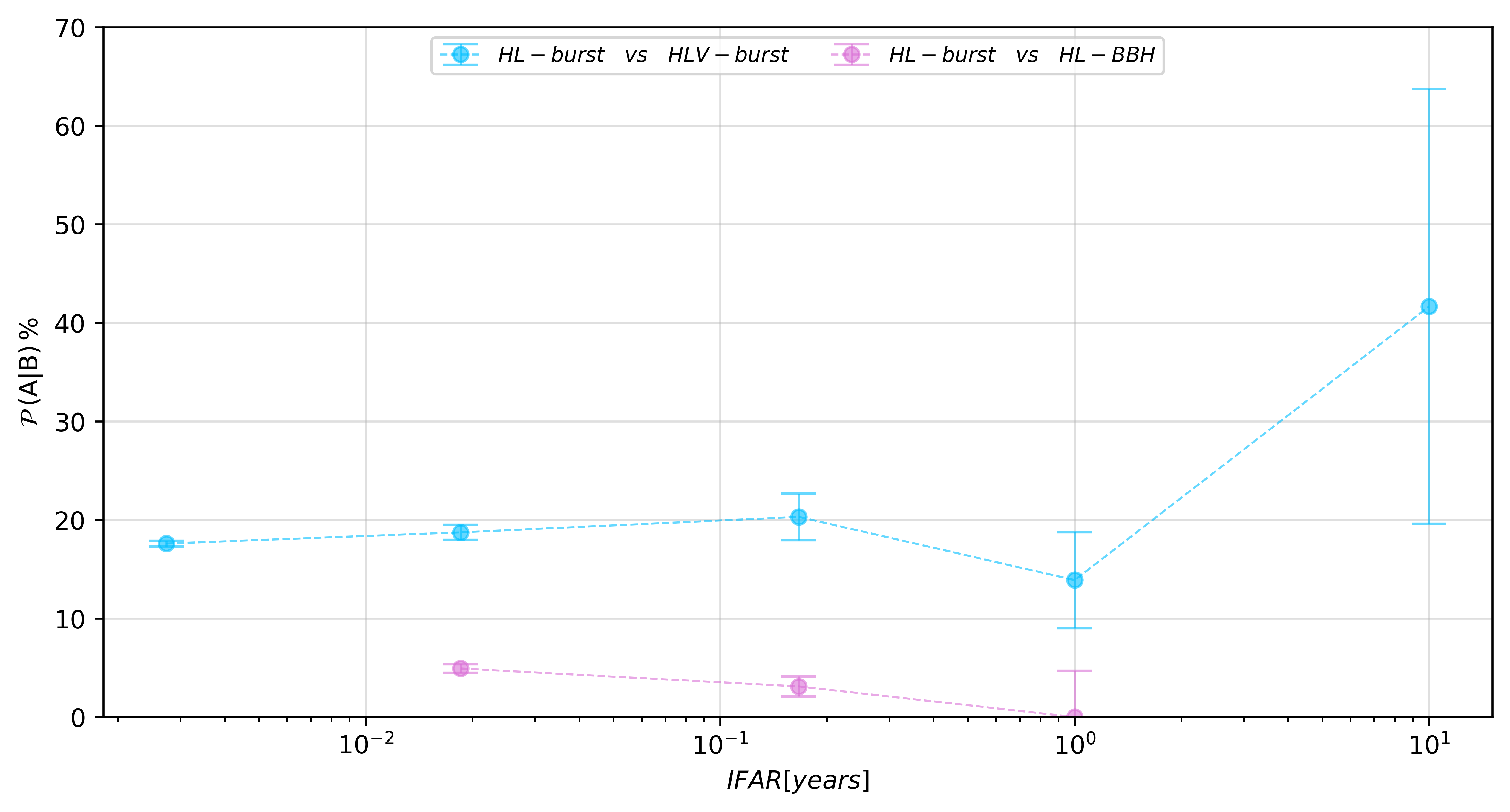}
\caption{Measured probabilities of common background triggers between concurrent searches HL-burst and HLV-burst (pale blue) and between HL-burst and HL-BBH (violet) as a function of the IFAR threshold.
In both cases, false alarms appear almost uncorrelated.
}
\label{fig:FA_conditioned_probability}
\end{figure}

The HL- and HLV-bursts show a small fraction of common false alarms; see Tab.\ref{tab:FAcomparison}. Moreover, the expected symmetry $\mathcal{P}(A | B)=\mathcal{P}(B | A)$ is empirically obeyed,  and the resulting probability of a common false alarm is $\mathcal{P}(B|A) \simeq 20\%$ for IFAR thresholds $\leq$ 1 year, see Fig. \ref{fig:FA_conditioned_probability}. 
While the statistical uncertainties limit the power of this investigation at higher IFAR thresholds, we can conservatively use a trials factor value equal to 2 for the combined search made by the logical OR of HL--burst and HLV--burst concurrently running at the same IFAR threshold. In fact, the measured trials factor $\simeq 1.8$ would only provide a $\simeq 10\%$ gain on the reported statistical significance of GWT candidates at IFAR $\leq$ 1 year. 

The comparison between HL-burst and HL-BBH leads to the same main conclusion about the trials factor. Here, the probability of common false alarms is even lower, limited to a few \%, see Fig. \ref{fig:FA_conditioned_probability}. This decorrelation is entirely due to the different ranking provided by XGBoost on the same set of transient candidates identified by cWB-2G on HL data. Here too, there is no evidence for a dependence of the probability of common false alarms on the statistical significance of the search, though the statistical uncertainties hinder this conclusion as our statistic vanishes at IFAR thresholds $>$ 1 year, see Tab.\ref{tab:FAcomparison}.



\section{Comparing sensitivities of different searches for GWTs}
\label{sec:comparison_genericvstargeted}
\label{sec:comparison}
In this section, we discuss the sensitivities of the searches on the same observing time from LVK O3 public data (see Sec. \ref{sec:dataset}). Their detection efficiency are compared at equal statistical significance, i.e. selecting equal IFAR thresholds, by analyzing the same set of software signal injections. This procedure enables the investigation of common and exclusive detections in different searches. In subsection \ref{subsec:HL_HLV_SIM}, we benchmark the sensitivity  of HL-burst and HLV-burst, using the visible volume as a proxy for the probability of detection of GWTs. Subsection \ref{subsec:HL_HLV_combined} discusses the performances of combined searches built from HL-burst and HLV-bursts. Finally,
subsection \ref{subsec:BBH_burst_SIM} compares results of the model--informed HL-BBH and the agnostic HL-burst. 

\subsection{Comparing searches for generic GWTs on HL and HLV data.}
\label{subsec:HL_HLV_SIM}

The performances of all--sky searches for short--duration bursts are compared by means of common signal injections from a set of ad-hoc waveforms, similar to the set used for testing the XGBoost classifier and described in Subsec. \ref{subsec:trainingset}. The signals include linearly polarized Gaussian pulses (GA), linearly polarized sine-Gaussian waveforms (SG), elliptically polarized sine-Gaussian waveforms (SGE) and white noise bursts with random polarization amplitudes (WNB). The polarization angle and, for SGEs, the inclination angle are randomly sampled. 

Here, differently from what previously done in \cite{Marek:2023}, we simulate a population of sources uniformly distributed in volume and radiating the same GW strength, i.e. with fixed root--squared--sum strain amplitude at some reference distance, namely $h_{rss}=10^{-22} /\sqrt{\mathrm{Hz}}$ at 1 Mpc. We consider this simple model as a suitable proxy for a wide class of plausible astrophysical populations of sources, respecting the generic criteria introduced in Subsec. \ref{subsec:trainingset}.\footnote{Given the general aims and wide unknowns of our GWT searches, we did not consider effects from the cosmological red-shift, inhomogeneities in the nearby Universe or variable emission strength in the source model.}

In this context, the simplest benchmark which tracks the expected rate of detections is the visible volume, measured as $V_{vis}(\text{IFAR}) = N_{detections}(\text{IFAR}) / \rho_{sources}$, where $N_{detections}(\text{IFAR})$ is the count of sources detected by the search above the IFAR threshold and $\rho_{sources}$ is the count of simulated sources per unit volume, set to the same value for each waveform model. The visible volume can be scaled easily in case of different choices of reference $h_{rss}$ amplitudes, as $V_{vis-hrss}(\text{IFAR}) = V_{vis}(\text{IFAR}) \ [ h_{rss} \sqrt{Hz} / 10^{-22} ]^3$.

\begin{figure}
\includegraphics[width=\textwidth]{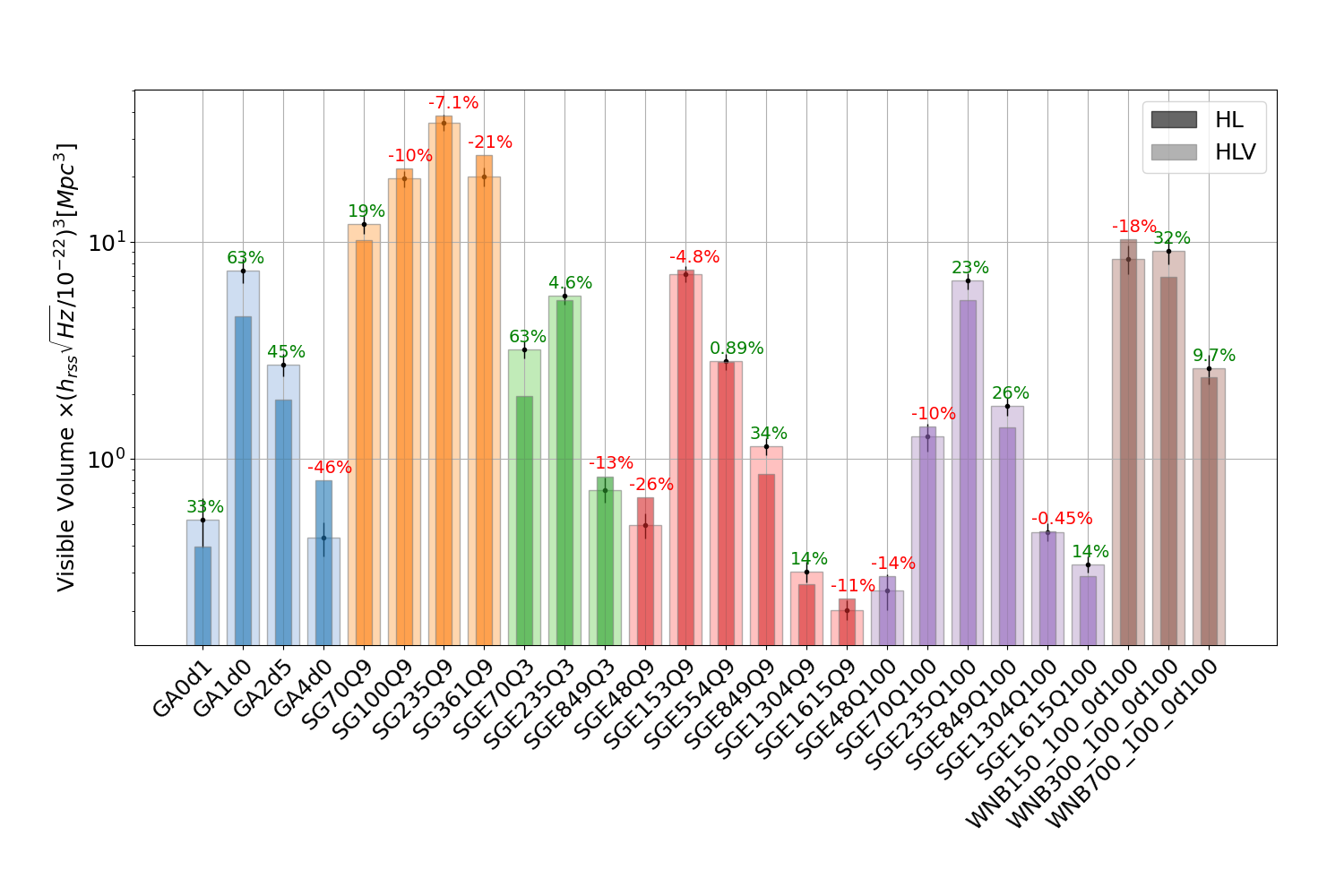}
\caption{Visible volume of HLV-burst (light color) and HL-burst (full color) at threshold IFAR=100 years on the same observing time from LVK O3. Sources are modeled as uniformly distributed in volume and emit at a fixed $h_{rss}$ different GW waveform models: linearly polarized Gaussian pulses (GA), linearly polarized sine--Gaussian (SG), elliptically polarized sine--Gaussian (SGE) and white noise burst (WNB). 
}
\label{fig:HL_HLV_hrss}
\end{figure}

Fig. \ref{fig:HL_HLV_hrss} reports the visible volumes achieved by HLV-burst and HL-burst at IFAR threshold equal to 100 years, $V_{vis}(\text{100y})$. At this high significance threshold, the two searches appear overall comparable: in half of the tested signal waveforms their visible volumes are within $\pm 15\%$ of each other, and HLV-burst performs better than HL-burst in 14 cases out of 26. 
This is the first demonstration that achieved visible volumes using latest public HLV data reach comparable values to those obtained using only the HL subset. 
Moreover, the overall performances of HL-burst reported in Fig. \ref{fig:HL_HLV_hrss} show a mean relative improvement in $V_{vis}$ of $\simeq$ +57\% with respect to what previously reported for cWB-2G on O3 HL data \cite{Marek:2023}, with a median value of +45\% and a variability between -30\% and +270\% on the 20 common waveform models. 
These progresses are indicative of the better performance of the new ranking procedure of candidates, fully based on the XGBoost classifier.

The variability of the visible volumes plotted in Fig. \ref{fig:HL_HLV_hrss} is evidently correlated to the frequency bandwidth content of each waveform and to its polarization model. The former effect is traced down to general properties of the noise spectra of the detectors. 
The effect of the polarization model is mediated by the directional sensitivity of the detectors and by the following convention, which impacts the signal strength on Earth. For GA, SG and WNB signals, the emission strength is modeled as isotropic and the conventional $h_{rss}$ is the actual root--squared--sum of the strain amplitude of the GWT at earth. Instead, SGE sources are modeled by the quadrupole emission formula and their conventional $h_{rss}$ is referred to the optimal direction of emission; the signal is then projected on Earth by a random inclination angle with the line of sight. This is the main reason for the larger visible volumes in case of SG signals with respect to SGE ones.

\begin{figure}
\includegraphics[width=0.95\textwidth]{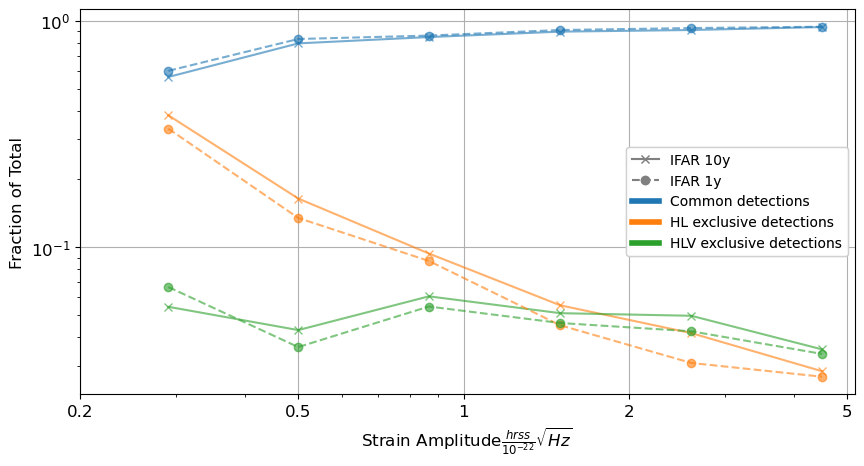}
\caption{Fraction of software signal injections recovered exclusively by HL-burst (orange), by HLV-burst (green) and by both analyses (blue) as a function of signal strain amplitudes $h_{rss}$. Two significance thresholds are considered: IFAR $>$ 10 yr (continuous curves) and $>$ 1 yr (dashed curves). The injections equally weight the signal models plotted in Fig.\ref{fig:HL_HLV_hrss}, but here the the source is imposing a grid of $h_{rss}$ amplitudes.}
\label{fig:HL_HLV_uniqueevents}
\end{figure}

A different Monte Carlo is used to investigate the complementarity and correlation of detectable signals in HL--burst and HLV--burst as a function of signal strength. Here, the same set of waveform models reported in Fig. \ref{fig:HL_HLV_hrss} are injected at a grid of selected $h_{rss}$ amplitude values.
Fig.\ref{fig:HL_HLV_uniqueevents} shows the fractions of common and exclusive detections as a function of the signal strength 
for IFAR thresholds of 1 year and 10 years, summing together the detections from an equal number of software injections per waveform model. 
As expected, the strongest signals tested, with $h_{rss} \geq 2 \times 10^{-22} /\sqrt{\mathrm{Hz}}$, are typically found by both HL-burst and HLV-burst, with only a few percent of the detections exclusively identified by only one search. In this regime, 
the detection efficiency is close to unitary for most waveform models, and HL-burst and HLV-burst appear of comparable power. 
At not--so--strong amplitudes, the fraction of common detections decreases and the exclusive contribution of HL-burst gets more important.
This asymmetry is consistent with previous reports of detection performances on O3 data \cite{allsky3, Marek:2023}. In fact, cWB-2G is constraining the search on HL data to the most favorable GW polarization component per each sky direction, while the exploitation of a detector of different directional sensitivity such as Virgo calls for considering both GW polarization components. The HL analyses then allow deploying stronger mitigation procedures of noise outliers with respect to HLV analyses. 
Overall, these results show only minor variations as a function of the search threshold in the IFAR range from 1 to 10 years.\footnote{Investigating the regime of lower signal amplitudes or higher IFAR thresholds is not technically feasible by this Monte Carlo. In fact, since the detection efficiency drops very rapidly with decreasing signal strength, it becomes computationally inefficient. The workaround which skips the analyses of too faint injections, which is deployed for the visible volume simulations, is not yet available here.} 

\begin{figure}
\begin{subfigure}[b]{0.328\textwidth}
        \centering
        \includegraphics[width=\textwidth]{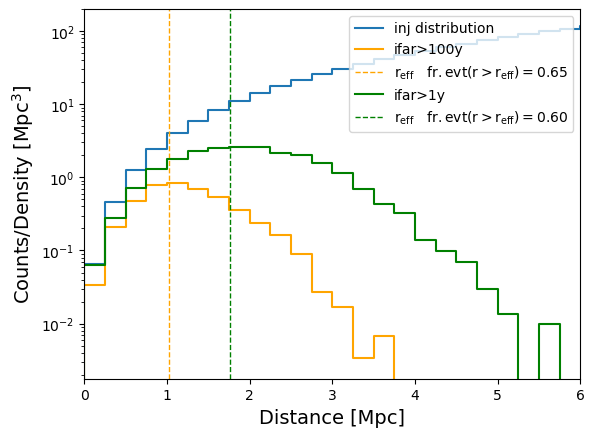}
        \caption{GA1d0}
        \label{fig:GA1d0}
\end{subfigure} 
\hfill
\begin{subfigure}[b]{0.328\textwidth}
        \centering
        \includegraphics[width=\textwidth]{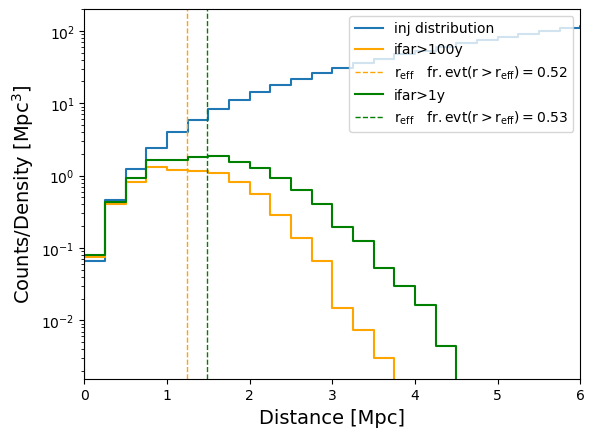}
        \caption{SGE153Q9}
        \label{fig:SGE153Q9}
\end{subfigure} 
\hfill
\begin{subfigure}[b]{0.328\textwidth}
     \centering
      \includegraphics[width=\textwidth]{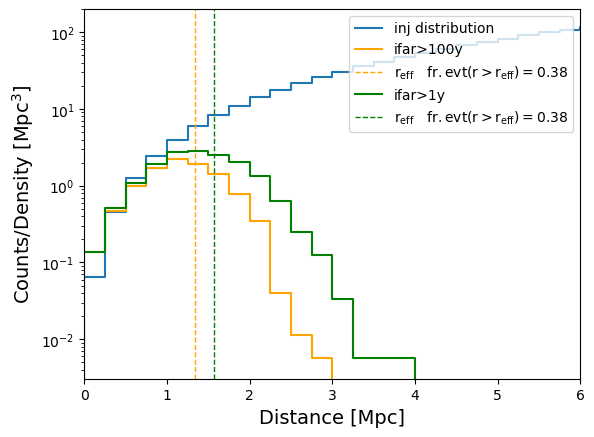}
        \caption{WNB150\_100\_0d1}
        \label{fig:WNB150_100_0d1}
\end{subfigure}
\caption{Visible volume of HL-burst as a function of source distance for three signal models, assuming a nominal strength $h_{rss}=10^{-22} \sqrt{Hz}$ at a reference distance of 1 Mpc: (a) Gaussian pulse of duration 1 ms (GA1d0); (b) sine-Gaussian elliptically polarized at 153 Hz with Q=9 (SGE153Q9); (c) white noise burst in the frequency band [150, 250] Hz with duration 0.1 s (WNB150\_100\_0d1).
The orange and green histograms report the increments of visible volume per bin in distance, at thresholds IFAR=100 y and 1 y respectively. The increments are estimated by the counts of detected sources belonging to the distance bin, divided by the density of simulated sources.
Vertical dashed lines indicate the radius of spheres whose volumes are equal to the visible volumes at the two thresholds. The blue stairway curves show the geometrical volume increment per distance bin, identical in all plots.}
\label{fig:visible_vol_distance}
\end{figure}

To investigate the dependence of the visible volume on the distance to the source, we increased the statistics of the Monte Carlo of sources uniformly distributed in volume and selected three waveform models: one GA, one SGE and one WNB. Fig.\ref{fig:visible_vol_distance} reports the increments of the visible volume as a function of distance to the source for the HL-burst search. The counts per waveform model are here increased by a factor $\sim$ 20 with respect to the simulation of Fig. \ref{fig:HL_HLV_hrss}. 
The increments of visible volume show an obvious general behavior common to the three signal models, i.e. the probability of detection is high at the shortest distances and vanishes farther away.
The behavior at larger distances depends on the signal model: in the case of randomly polarized WNBs the contributions to the visible volume steeply drop with distance, while for GA and SGE models the tails are more extended. We benchmark these effects by counting the fraction of detected signals beyond an effective radius defined as $[3/(4\pi) \ V_{vis}]^{1/3}$, where $V_{vis}$ is the visible volume for the waveform model considered. These fractions are $\simeq$ 0.63, 0.53 and 0.38 in case of GA, SGE and WNB respectively, with a minor dependence on the search threshold between IFAR = 1 yr and 100 yrs. In other words, detectable sources belong to a more compact volume for WNBs, while for GAs detectable sources are more dispersed. This effect is expected, since both GW polarization components are on average equally weighted in WNBs, so that the projection of these signals into the HL observatory defines a sharper horizon along any incident direction. At the other extreme, Gaussian pulses have a random linear polarization in the wavefront plane, which causes a larger variability of the signal projection on the HL observatory along any incident direction, resulting in a more blurred horizon.

Comparing results at IFAR thresholds of 1 yr and 100 yrs, the visible volume shrinks by a factor $\simeq$ 1.7 for the SGE and WNB models. Instead, for the Gaussian pulses the shrinking factor is much stronger, $\simeq$ 5.1. This is related to the similarity of GAs to the dominating family of stronger glitches in the LIGO detectors' data. In fact, the XGBoost training is tuned to the features of the background triggers, see \ref{subsec:bkg}, and glitch--like candidates are more frequently prevented from reaching a high rank than other waveform models.

\subsection{Comparing combined searches for generic GWTs on HL and HLV data.}
\label{subsec:HL_HLV_combined}
We address here the question of what is the best performing short--duration burst search over a given observing time, when data are available from three detectors. We focus in particular on concurrent HL-burst and HLV-burst analyses set at the same IFAR threshold, and on the combined searches made by the logical OR ($\lor$) and logical AND ($\land$) of their candidates, respectively HL$\lor$HLV-burst and HL$\land$HLV-burst. We exploit here the same set of uniformly distributed sources used in the investigation of the visible volume as a function of distance, Fig.\ref{fig:visible_vol_distance}. With this signal set, whose amplitude distribution is $\propto h_{rss}^{-2}$, HL-burst is detecting more signals than HLV-burst for IFAR thresholds $<$ 100 years, in agreement with the fact that HL-burst performs better than HLV-burst for signal amplitudes in the lower amplitude range plotted in 
Fig.\ref{fig:HL_HLV_uniqueevents}. Taking HL-burst as reference, the fraction of signals that are common detections between HL-burst and HLV-burst is rather flat around $\sim$ 92\% over most IFAR thresholds, for all the three waveform models tested. On the other side, the union of HL-burst and HLV-burst detections surpasses the HL-burst detections by an almost constant relative increment, $\sim$ 10\%.

\begin{figure}
\begin{subfigure}[b]{0.328\textwidth}
        \centering
        \includegraphics[width=\textwidth]{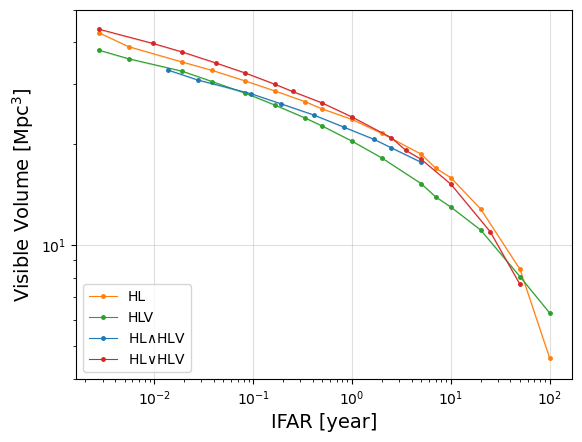}
        \caption{GA1d0}
        \label{fig:GA1d0}
\end{subfigure} 
\hfill
\begin{subfigure}[b]{0.328\textwidth}
        \centering
        \includegraphics[width=\textwidth]{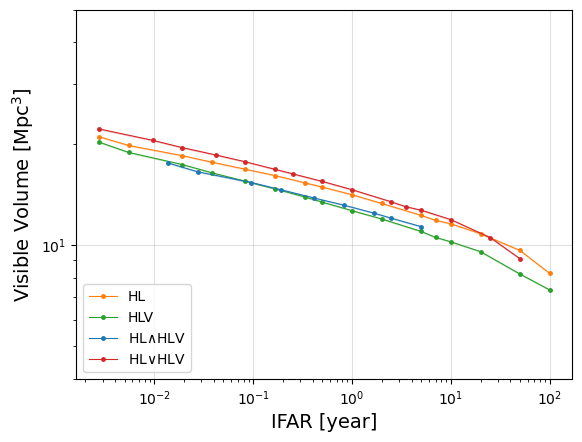}
        \caption{SGE153Q9}
        \label{fig:SGE153Q9}
\end{subfigure} 
\hfill
\begin{subfigure}[b]{0.328\textwidth}
     \centering
      \includegraphics[width=\textwidth]{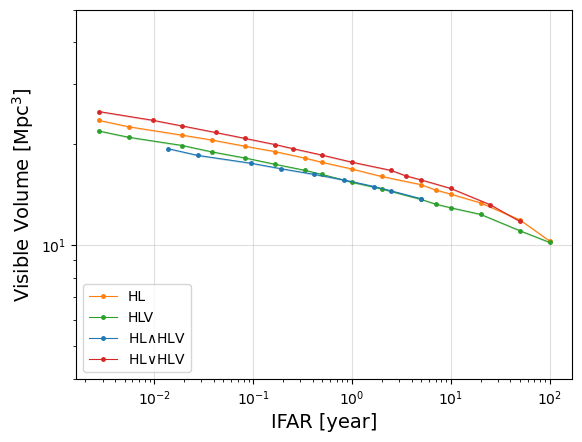}
        \caption{WNB150\_100\_0d1}
        \label{fig:WNB150_100_0d1}
\end{subfigure}
\caption{Comparison of visible volumes achieved by different searches as a function of the IFAR threshold, for the waveform models GA1d0 (a), SGE153Q9 (b) and $\text{WNB150\_100\_0d1}$ (c).
The colored curves and belts show visible volumes and one $\sigma$ statistical uncertainties for HL-burst (orange), HLV-burst (green), HL$\lor$HLV-burst (red) and HL$\land$HLV-burst (blue).
}
\label{fig:visiblevolumecomparison}
\end{figure}

The comparison of visible volumes as a function of the IFAR threshold is shown in Fig.\ref{fig:visiblevolumecomparison}. 
The visible volume of HL$\lor$HLV-burst is estimated by counting the union of the detected signal sets from HL-burst and HLV-burst analyses and by halving the IFAR of HL-burst and HLV-burst. 
In fact, these analyses show mostly independent false alarms on O3 data, and the use of a trials factor of 2 brings a small $\simeq$ 10\% bias on the conservative side (see Sec.\ref{sec:FAcorrelations}). 
For the first time, we are demonstrating a systematic gain in visible volume of HL$\lor$HLV-burst with respect to HL-burst for IFAR thresholds in a wide range, up to $\sim$ 10 years. 
The relative gain in probability of detection on O3 data is up to $\sim$ 8\% in the three signal models investigated, and decreases as the IFAR threshold increases. The threshold range where HL$\lor$HLV-burst surpasses HL-burst is narrower in the case of the waveform model GA. This is consistent with the fact that the visible volumes for GA signals are more rapidly decreasing with IFAR than for other waveforms. 

The visible volume of HL$\land$HLV-burst is estimated by counting the common detections of HL-burst and HLV-burst from the same Monte Carlo. The statistical significance of HL$\land$HLV-burst is instead higher by a factor of $\simeq 5$ in IFAR with respect to the original HL-burst and HLV-burst, up to a significance of $\sim$ 5 years.\footnote{The false alarm statistics available is insufficient to estimate a useful lower limit to the resulting IFAR of HL$\land$HLV-burst for higher search thresholds, see Fig.\ref{fig:FA_conditioned_probability}.} 
As shown in Fig.\ref{fig:visiblevolumecomparison}, the resulting visible volumes of HL$\land$HLV-burst are not competitive with respect to the performances of HL-burst in the plotted range of thresholds.

\subsection{Comparing sensitivities of agnostic and BBH--model--informed searches.}
\label{subsec:BBH_burst_SIM}

The model-informed search HL-BBH is midway between generic searches, that do not assume a signal model, and template-based searches, which rely on accurate and extensive template banks. As described in subsec.\ref{subsec:trainingset}, our HL-BBH utilizes the same cWB-2G data processing and same candidate triggers of HL-burst, while it exploits a specific training of the XGBoost classifier that targets compact binary coalescences using a more extended set of morphological features of triggers. 
In this case, HL-BBH can be deployed on top of the HL-burst with a negligible addition of computational resources, while delivering better detection performances for BBH and IMBH sources with respect to HL-burst. However, the performances of this HL-BBH are suffering from the model agnostic preselection of triggers, which favors candidates with compact time--frequency representations and is far from optimal in the case of lower $\text{chirp mass}$ signals.  
There are recent examples of more advanced model--informed searches based on cWB targeting CBCs \cite{Mishra:2021tmu, Mishra:2024zzs} or hyperbolic encounters of compact objects \cite{Bini:2023gaj}.

\begin{figure}
\centering
\begin{minipage}{.5\textwidth}
  \centering
  \includegraphics[width=.99\linewidth]{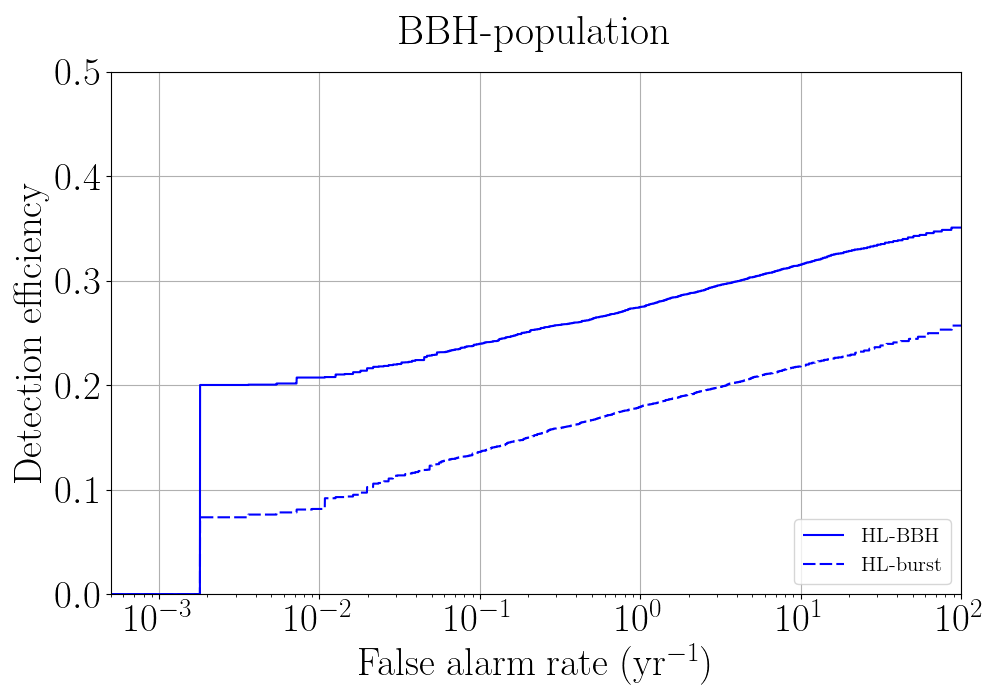}
\end{minipage}%
\begin{minipage}{.5\textwidth}
  \centering
  \includegraphics[width=.99\linewidth]{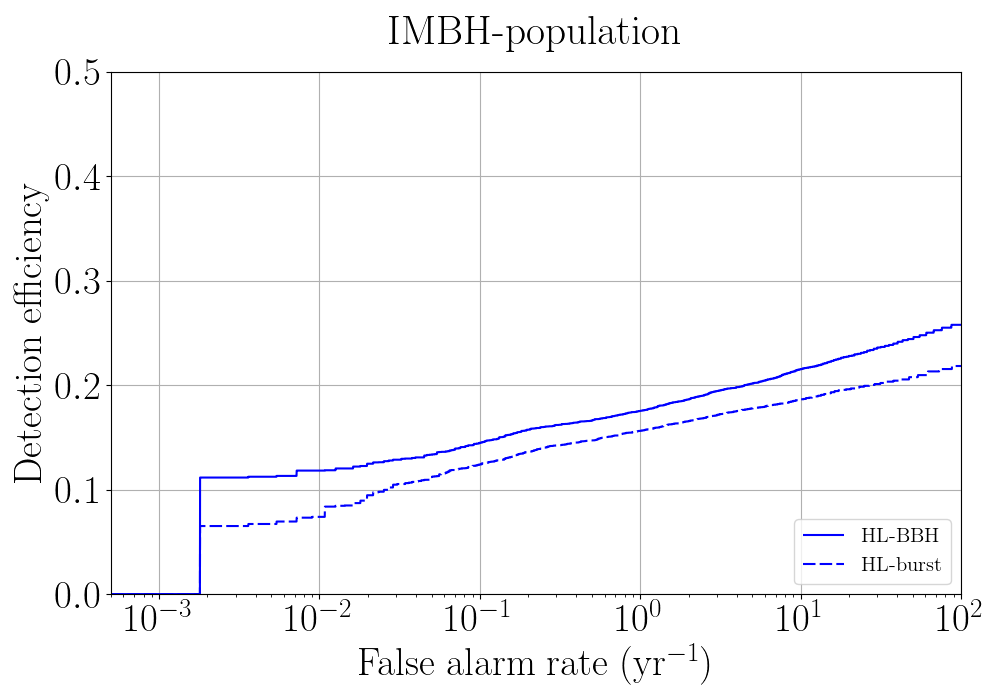}
\end{minipage}
\caption{Detection efficiency versus IFAR for an astrophysical population of CBC sources of stellar mass (left) and for the subset of IMBH sources (right) using the unmodeled search HL-burst (dotted line) and the model-informed search HL-BBH (continuous line). For the population model used see end of Subsec.\ref{subsec:trainingset}.}
\label{fig:generic_modelinf}
\end{figure}

Fig. \ref{fig:generic_modelinf} compares the receiver operating curves of the model--informed and model--agnostic searches over the stellar-mass CBC and IMBH population models described at the end of Subsec.\ref{subsec:trainingset}: the advantage in detection efficiency of HL-BBH is evident in the entire false alarm rate range. In particular, the probability of detection of stellar-mass CBC sources is approximately doubled with respect to HL-burst for candidate significances IFAR between 10 and 500 years, while the systematic improvement for IMBH signals is relevant but less striking, since HL-burst is already efficient for short--duration GWTs.
However, this is accomplished at the cost of highly constraining the detectable signal parameter's space of HL-BBH: the visible volumes achieved by HL-BBH for all the signal morphologies shown in Fig. \ref{fig:HL_HLV_hrss} become negligible with respect to what HL-burst achieves. This is expected, since estimated signal features such as central frequency, frequency bandwidth, duration, chirp mass and similarity to chirping signal morphology are used by HL-BBH for the multivariate classification of events.

\section{Summary and conclusions}
This work is a status update of the 2G version of the coherent WaveBurst pipeline, one of the pipelines used by the LVK Collaboration to search for generic GWTs in the current observing run (O4).
All the results discussed here are based on the analyses of 14.8 days of Hanford-Livingston-Virgo data released by LVK within the Observing Run 3, the most recent publicly available three--detector data. 
The main innovations are implemented in the post-processing stage of cWB-2G, where the ranking of candidate triggers is now entirely performed by the eXtreme Gradient Boosting classifier. This enhances the flexibility of the scope of cWB-2G, since its training more easily adapts to the characteristics of the detectors' network, of the noise triggers and of the target signal population. In particular, we focus on three searches: two targeting generic GWTs and using HL and HLV data, one targeting BBH and IMBH mergers using HL data.

We  
implement enhanced tools for validating the statistical significance of GWT candidates assessed by means of time-slides over a broad range of thresholds, at Inverse False Alarm Rates from 1 day to 1 century.
Moreover, we measure the correlation and complementarity of background triggers and simulated GWTs across different searches. These methods enable us to provide for the first time a comprehensive comparison of searches for generic GWTs on Hanford-Livingston and Hanford-Livingston-Virgo data. 

The most directly applicable results are reported in Sec. \ref{sec:comparison}, including benchmarks of the sensitivity to populations of sources and insights on the probability of detection as a function of distance to the source. 
For what concerns searches for short duration GWTs which run concurrently on the same observing time, the current cWB-2G version is capable to reach comparable performances when using HL or HLV data from O3 at high significance thresholds. Moreover, we demonstrate for the first time that the logical OR combination of the two concurrent searches provides a gain in probability of detection of short duration GWTs over a wide range of IFAR thresholds, improving visible volumes up to $\sim$ 8\% with respect to that exploiting only HL data at the same IFAR, which was 
the search previously setting the best performances. 
In addition, we demonstrate how the search can be model--informed by acting on the XGBoost classifier, and we compare a search for binary black holes of stellar and intermediate mass with a search for generic GWTs.

Based on this work, we shall design advantageous combinations of all-sky searches for GWTs provided by concurrent analyses, either unmodelled or weakly-modeled, on different combinations of GW detectors. In particular, our results provide important insights for assessing data analysis strategies in searches for short--duration GWTs in the ongoing LVK observing run, O4, and in future observations.
The methods discussed here have been demonstrated on the open source cWB-2G pipeline with XGBoost post-processing, but in principle they can be generalized to compare and combine all-sky searches for generic GWTs based on different pipelines.

We plan to extend this work on all-sky all-times GWT surveys to searches for targeted sources which leverage on additional information on their sky positions and GWT arrival times. This development will call for investigating different weakly-modeled approaches, as well as comparing and combining results from concurrent analyses on different detectors' configurations. For what concerns programming code developments, we are developing a Python-based implementation of the cWB-2G pipeline, PyCWB \cite{Xu:2023ioz, pycwb_repo}, which promises to easy further methodological progresses and usability by a broader research community.



\section*{Acknowledgments}
The authors are grateful to Dr. Leigh Smith for helpful discussions and feedback on this manuscript.

This research has made use of data or software obtained from the Gravitational Wave Open Science Center (gwosc.org), a service of the LIGO Scientific Collaboration, the Virgo Collaboration, and KAGRA. This material is based upon work supported by NSF's LIGO Laboratory which is a major facility fully funded by the National Science Foundation, as well as the Science and Technology Facilities Council (STFC) of the United Kingdom, the Max-Planck-Society (MPS), and the State of Niedersachsen/Germany for support of the construction of Advanced LIGO and construction and operation of the GEO600 detector. Additional support for Advanced LIGO was provided by the Australian Research Council. Virgo is funded, through the European Gravitational Observatory (EGO), by the French Centre National de Recherche Scientifique (CNRS), the Italian Istituto Nazionale di Fisica Nucleare (INFN) and the Dutch Nikhef, with contributions by institutions from Belgium, Germany, Greece, Hungary, Ireland, Japan, Monaco, Poland, Portugal, Spain. KAGRA is supported by Ministry of Education, Culture, Sports, Science and Technology (MEXT), Japan Society for the Promotion of Science (JSPS) in Japan; National Research Foundation (NRF) and Ministry of Science and ICT (MSIT) in Korea; Academia Sinica (AS) and National Science and Technology Council (NSTC) in Taiwan.

The authors are grateful for computational resources provided by the LIGO Laboratory, supported by National Science Foundation Grants PHY-0757058 and PHY-0823459. 

This work has been funded by the European Union-Next Generation EU, Mission 4 Component 1 CUP J53D23001550006 with the PRIN Project No. 202275HT58. This work is partially supported by ICSC – Centro Nazionale di Ricerca in High Performance Computing, Big Data and Quantum Computing, funded by European Union – NextGenerationEU.

MD acknowledges the support of the Sapienza Grants No. RM123188F3F2172A and RG1241905E47F60C. OGF acknowledges support by the Spanish Agencia Estatal de Investigaci\textbackslash{}'on (grants PID2021-125485NB-C21 and PID2024-159689NB-C21) funded by MCIN/ AEI/10.13039/501100011033 and ERDF A way of making Europe, as well as by the Portuguese Foundation for Science and Technology (FCT) through doctoral scholarship UI/BD/154358/2022, and CF-UM-UP through Strategic Funding UIDB/04650/2020.

\appendix

\section{Configuration of analyses}
\label{app:config}
\subsection{tuning of XGBoost hyperparameters.}
\label{subsec:tuningXGBoost}
In this work, we utilize the XGBoost classifier \cite{XGBoost}, 
whose structure is defined by hyperparameters. 
The tuning of the hyperparameters is performed on a validation dataset 
using a K-fold cross validation: the original dataset is randomly shuffled and split into k different subsets, of which K-1 are used for training the model, and the remaining one is used to validate the performance.
The procedure is repeated K times, looping over the subsets, so that at every iteration, a different subset is treated as an independent validation set. The evaluation results are computed and stored at every iteration, and they are combined at the end to provide a final, quantitative assessment of the model's performance.
Hyperparameters' tuning involves evaluating performance metrics such as the receiver operating characteristic (ROC) curve and the precision-recall (PR) curve. In particular, we considered the areas under the PR curve (PR-AUC) and ROC curve (ROC-AUC). The former is particularly relevant for imbalanced datasets, since it benchmarks the trade-off between false positives and the rate of false negatives. In addition, we considered a particular version of the ROC-AUC metric built from the logarithmic values of detection efficiency and false alarm probability. In fact, this enhances the benchmark's sensitivity to the region of confident detections, i.e. in the lower range of investigated false alarm probabilities.  

The tuning of XGBoost's hyperparameters has been performed manually and separately for the short duration burst searches from the HL-BBH, which use different sets of trigger features (see next subsection). The selected values of hyperparameters are reported in in Tab. \ref{tab:hyperparameters}. We checked that the performances of the searches were not crucially depending on plausible changes with respect to the tuned values.

\begin{table}[h]
\centering
\begin{tabular}{ l | c c c } 
\hline
\textbf{Parameter} & \textbf{range} & \textbf{HL and HLV-burst} & \textbf{HL-BBH} \\
\hline
learning\_rate & (0,1) &  0.003 & 0.03 \\ 
max\_depth & (0,$\infty$) & 8 & 13 \\  
min\_child\_weight & (0,$\infty$) & 3 & 2 \\  
colsample\_bytree & (0,1] & 0.5 & 0.7 \\  
subsample & (0,1] & 0.8 & 0.6 \\  
gamma & [0,$\infty$] & 0.4 & 1.0 \\
\hline
\end{tabular}
\caption{List of XGBoost hyperparameters (first column), their allowed range (second column), the tuned values for HL-burst and HLV burst (third column), and the tuned values for HL-BBH (fourth column).}
\label{tab:hyperparameters}
\end{table}

Here, \textbf{learning\_rate} regulates how much each generated decision tree affects the successive one by shrinking the weights of the features (the lower the less prone to overfitting).
The \textbf{max\_depth} parameter determines the deepness of each decision tree, deeper trees can capture more complex patterns in the data, but may also lead to overfitting.
The other parameters \textbf{min\_child\_weight, colsample\_bytree, subsample, and gamma} act as different forms of regularization to preserve the conservativeness of the algorithm and prevent overfitting \cite{XGBoost,cwb_manual}.

\subsection{XGBoost training procedure}
\label{subsec:trainingXGBoost}
Training the XGBoost ML algorithm provides a decision model capable of discriminating between GWT triggers and background triggers.
The background triggers are sampled by analyzing time-slides of the original data, 
that provide an off-source data set which is typically more than $10^5$ times longer than the on-source observing time. 
A Monte Carlo of GWT injections provides the set representing genuine signals. 

These data sets are split in two separate halves, one half is used for training XGBoost, while the second half is used to test the trained model by measuring the statistical significance of candidates and the detection efficiency for the simulated population of GWTs. Further alternative population models of GWTs are simulated to interpret the search under different conditions. 

\begin{table}
\centering
\renewcommand{\arraystretch}{1.2} 
\renewcommand{\baselinestretch}{0.8} 
\begin{tabular}{l|p{11cm}} 
\hline
Feature & Description \\ 
\hline
norm            & sum of energies appearing in all TF resolution layers divided by energy of the event. Lower values mean the signal energy is concentrated in very few TF layers, a characteristic which is very common for background triggers. \\
noise           & $\sqrt{\frac{1}{\sum_{\mathrm{det}}\frac{1}{N_{\mathrm{det}}}}} $ where $N_{\mathrm{det}}$ is the noise energy in each detector. \\
$c_c$           & network correlation coefficient. \\
$\chi^{2}$      & noise energy per degree of freedom. \\
mSNR/likelihood & max value of $\mathrm{SNR}^2$ in single detectors divided by the total likelihood in the network. \\
ecor/likelihood & sum of the off-diagonal term of likelihood matrix normalized to the total likelihood.\\
rho0\_100d0     & effective correlated SNR, capped at 100.\\
Qa              & similarity of waveform to one dominant pulse, inspired to blip glitches.\\
Qp              & effective number of oscillations in the waveform. \\                                                                                       \hline                                                                                         
\end{tabular}
\caption{List of the trigger's features reconstructed by cWB-2G and passed to XGBoost for HL-burst and HLV-burst.}
\label{tab:xgboost_features}
\end{table}

\begin{table}
\centering
\renewcommand{\arraystretch}{1.2} 
\renewcommand{\baselinestretch}{0.8} 
\begin{tabular}{l|p{11cm}} 
\hline
Feature & Description \\ 
\hline
norm            & same as in burst searches. \\
$c_c$           & same as in burst searches. \\ 
$\chi^{2}$      & same as in burst searches. \\ 
sSNR/likelihood & $\mathrm{SNR}^2$ of L detector divided by the total likelihood in the network. \\
ecor/likelihood & same as in burst searches.\\
rho0\_25d0     & effective correlated SNR, capped at 25.\\
Qa              & same as in burst searches.\\
Qp              & same as burst searches. \\
frequency       & mean frequency of the recovered $\mathrm{SNR}^2$.\\
duration    &  standard deviation of the time distribution of the recovered  $\mathrm{SNR}^2$.\\
bandwidth    &  standard deviation of the frequency distribution of the recovered  $\mathrm{SNR}^2$.\\
mchirp       & estimate of chirpmass \cite{V.Tiwari_2016}.\\
mchirp3-5      & three different metrics of the goodness-of-fit of the chirpmass estimate \\

\hline          
\end{tabular}
\caption{List of the trigger's features reconstructed by cWB-2G and passed to XGBoost for HL-BBH.}
\label{tab:xgboost_features_BBH}
\end{table}

The 9 trigger's features that cWB-2G passes to XGBoost for the HL-burst and HLV-burst are described in Tab. \ref{tab:xgboost_features}. HL-BBH instead uses the 15 features listed in Tab. \ref{tab:xgboost_features_BBH}.


\section{Poisson checks and Background stability}
\label{app:BKG}
\begin{figure}
\centering
\includegraphics[width=0.8\textwidth]{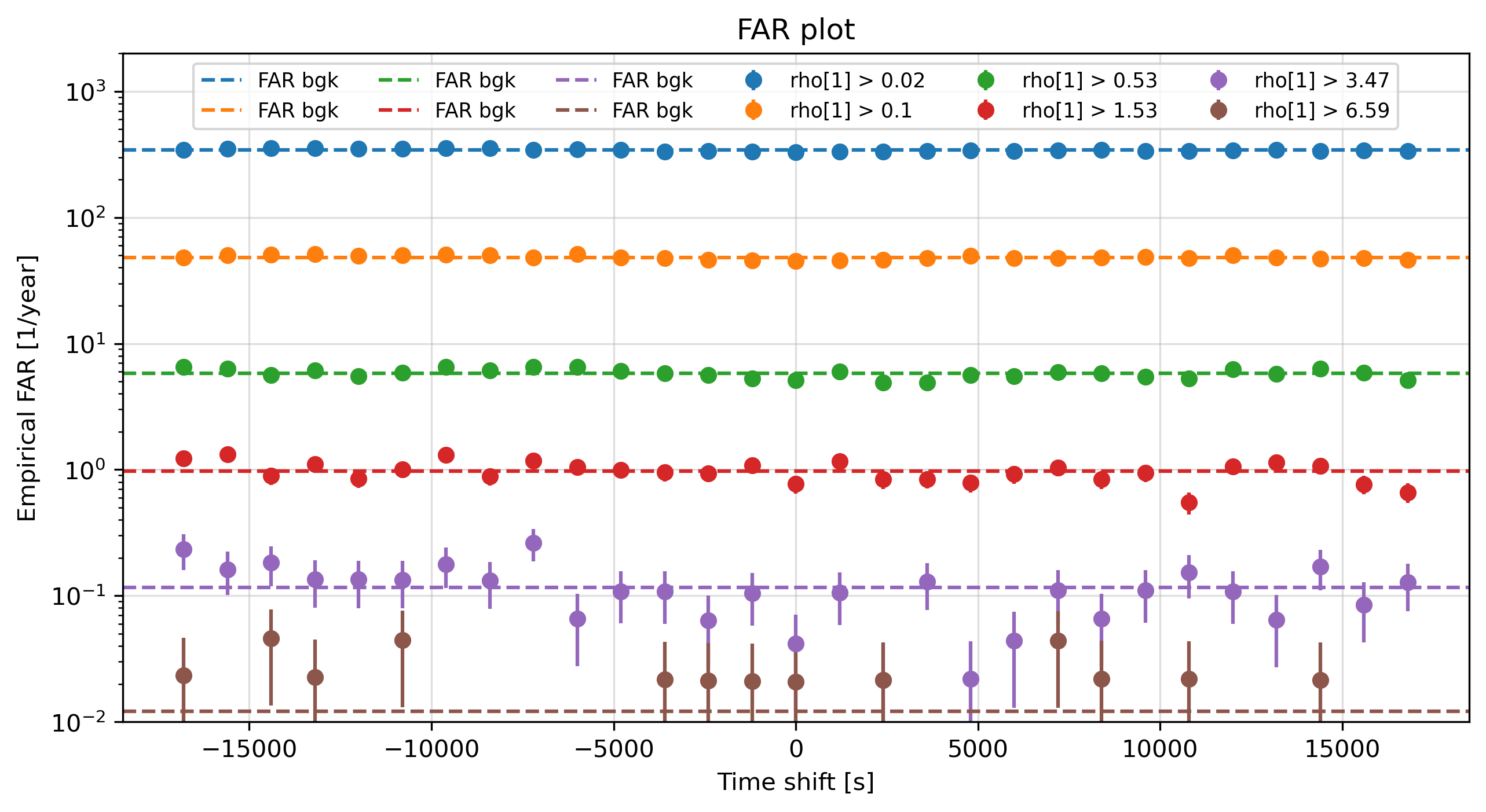}
\caption{FAR values at fixed rho thresholds as a function of the average shift in time of disjoint sets of time slides. The time shift is applied to the Hanford data stream.}
\label{fig:FARslags}
\end{figure}

Here we report the details of the study discussed in section \ref{sec:significance} of this paper.
The following tables show the empirical estimate of the FAR ($\text{FAR}^{\text{slag}}_{\text{emp}}$) for each segment-shift and $\eta^{\prime}_0$ thresholds ($\{\text{A}, \text{B}, \text{C}, \text{D}, \text{E}, \text{F}\}$) for the tree different searches: HL-burst (TAble \ref{tab:FARslagsLH}), HLV-burst (Table \ref{tab:FARslagsLHV}), and HL-BBH (Table \ref{tab:FARslagsBBH}) respectively.
Each $\text{FAR}^{\text{slag}}_{\text{emp}}$ is provided with its relative 95$\%$ credible interval (in square brackets) and in bold are highlighted the  $\text{FAR}^{\text{slag}}_{\text{emp}}$ values which fall off this credible interval.
Then, $\sigma[\text{FAR}^{\text{slag}}_{\text{emp}}]$ and $\sigma[\text{FAR}^{\text{slag}}_{\text{th}}]$ provide the variance of the $\text{FAR}^{\text{slag}}_{\text{emp / th}}$ set for each $\eta^{\prime}_0$ threshold value, and are referred to the empirical (emp.) FAR estimate and theoretical (th.) one, so the one recovered from the Poissonian estimate $\mathcal{P}(k;\mu)$.
The last two rows of each table provide the empirical and theoretical FAR ($\text{FAR}^{\text{bkg}}_{\text{emp}}$ and $\text{FAR}^{\text{bkg}}_{\text{th}}$ respectively) estimate for the entire background with their uncertainty.
We can see that when dealing with $\eta^{\prime}_0$ thresholds corresponding to FAR smaller than one per year then $\text{FAR}^{\text{slag}}_{\text{th}}$ result to be smaller or comparable to $\text{FAR}^{\text{slag}}_{\text{emp}}$.
This behaviour points to an expected weakness of the XGBoost model in rejecting possible noise realisation with morphologies too similar to the signals used in the training set (see section \ref{sec:dataset}).
Nevertheless, this systematic uncertainty excess is limited to few $\%$ with no impact on the analysis, as it is possible to see from the estimate of $\text{FAR}^{\text{bkg}}_{\text{emp}}$ and $\text{FAR}^{\text{bkg}}_{\text{th}}$. 
Opposite, we can see that for $\eta^{\prime}_0$ thresholds corresponding to FAR greater than one per year we fall into a possible cooling of the background.
This would suggest that XGBoost rejection capability, when dealing with such rare noise outliers, performs so well that the number of detected events is slightly smaller than its Poissonian prediction.
Nevertheless, these are speculations since Figure \ref{fig:FAR_rms}, of section \ref{sec:significance} shows that these fluctuations of $\sigma[\text{FAR}^{\text{slag}}_{\text{emp / th}}]$ are compatible within their uncertainties.

Figure \ref{fig:FARslags} shows, for HL-burst, that the empirical FAR estimate for each segment-shift, $\text{FAR}^{\text{slag}}_{\text{emp}}$,  does not display any trend or estimate bias, pointing to robustness in background estimation through the procedure of time-shifts.
This is a visual representation of the information encapsulated in table \ref{tab:FARslagsLH}.

\begin{table}
\footnotesize
\lineup
\renewcommand{\arraystretch}{1.2}
\begin{tabular}{@{}*{7}{c}}

\br                              
 & $\text{A}_{\text{ LH}}$ & $\text{B}_{\text{ LH}}$  & $\text{C}_{\text{ LH}}$ &  $\text{D}_{\text{ LH}}$ & $\text{E}_{\text{ LH}}$ & $\text{F}_{\text{ LH}}$ \cr

\mr
\multirow{2}{*}{Time shift [s]} &
\multicolumn{6}{c}{$ \text{FAR}^{\text{slag}}_{\text{emp}} \quad {\scriptscriptstyle [\text{FAR}_{\text{slag, th}}^{5\%}, \, \text{FAR}_{\text{slag, th}}^{95\%}] } $} \\
 & {\tiny [1/day]} & {\tiny [1/week]} & {\tiny [1/6 month]} & {\tiny [1/year]} & {\tiny [1/10years]} & {\tiny [1/100years]} \\

\mr
-16800 & $1.0 \quad {\scriptscriptstyle [0.99,1.01] }$ & $0.98 \quad {\scriptscriptstyle [0.97,1.04] }$ & $1.1 \quad {\scriptscriptstyle [0.8,1.2] }$ & $1.1 \quad {\scriptscriptstyle [0.7,1.2] }$ & $1.4 \quad {\scriptscriptstyle [0.2,1.9] }$ & $2.3 \quad {\scriptscriptstyle [0.0,4.7] }$ \cr       
-15600 & $1.01 \quad {\scriptscriptstyle [0.99,1.01] }$ & $1.02 \quad {\scriptscriptstyle [0.97,1.04] }$ & $1.1 \quad {\scriptscriptstyle [0.8,1.2] }$ & $\mathbf{ 1.3 } \quad {\scriptscriptstyle [0.8,1.2] }$ & $1.4 \quad {\scriptscriptstyle [0.2,1.9] }$ & $0.0 \quad {\scriptscriptstyle [0.0,4.6] }$ \cr       
-14400 & $\mathbf{ 1.02 } \quad {\scriptscriptstyle [0.99,1.01] }$ & $1.02 \quad {\scriptscriptstyle [0.97,1.04] }$ & $0.9 \quad {\scriptscriptstyle [0.8,1.2] }$ & $0.8 \quad {\scriptscriptstyle [0.7,1.2] }$ & $1.4 \quad {\scriptscriptstyle [0.2,1.8] }$ & $2.3 \quad {\scriptscriptstyle [0.0,4.6] }$ \cr       
-13200 & $\mathbf{ 1.03 } \quad {\scriptscriptstyle [0.99,1.01] }$ & $\mathbf{ 1.04 } \quad {\scriptscriptstyle [0.97,1.03] }$ & $1.1 \quad {\scriptscriptstyle [0.8,1.2] }$ & $1.0 \quad {\scriptscriptstyle [0.8,1.2] }$ & $1.1 \quad {\scriptscriptstyle [0.2,1.8] }$ & $2.3 \quad {\scriptscriptstyle [0.0,4.5] }$ \cr       
-12000 & $\mathbf{ 1.02 } \quad {\scriptscriptstyle [0.99,1.01] }$ & $1.02 \quad {\scriptscriptstyle [0.97,1.03] }$ & $\mathbf{ 0.8 } \quad {\scriptscriptstyle [0.8,1.2] }$ & $0.8 \quad {\scriptscriptstyle [0.7,1.2] }$ & $0.7 \quad {\scriptscriptstyle [0.2,1.8] }$ & $0.0 \quad {\scriptscriptstyle [0.0,4.5] }$ \cr       
-10800 & $\mathbf{ 1.02 } \quad {\scriptscriptstyle [0.99,1.01] }$ & $1.02 \quad {\scriptscriptstyle [0.97,1.03] }$ & $1.0 \quad {\scriptscriptstyle [0.8,1.2] }$ & $0.9 \quad {\scriptscriptstyle [0.7,1.2] }$ & $0.9 \quad {\scriptscriptstyle [0.2,1.8] }$ & $2.2 \quad {\scriptscriptstyle [0.0,4.5] }$ \cr       
-9600 & $\mathbf{ 1.02 } \quad {\scriptscriptstyle [0.99,1.01] }$ & $\mathbf{ 1.04 } \quad {\scriptscriptstyle [0.97,1.03] }$ & $1.2 \quad {\scriptscriptstyle [0.8,1.2] }$ & $1.2 \quad {\scriptscriptstyle [0.7,1.2] }$ & $1.1 \quad {\scriptscriptstyle [0.2,1.8] }$ & $0.0 \quad {\scriptscriptstyle [0.0,4.5] }$ \cr       
-8400 & $\mathbf{ 1.02 } \quad {\scriptscriptstyle [0.99,1.01] }$ & $1.03 \quad {\scriptscriptstyle [0.97,1.04] }$ & $0.9 \quad {\scriptscriptstyle [0.8,1.2] }$ & $0.8 \quad {\scriptscriptstyle [0.7,1.2] }$ & $0.9 \quad {\scriptscriptstyle [0.2,1.8] }$ & $0.0 \quad {\scriptscriptstyle [0.0,4.4] }$ \cr       
-7200 & $0.99 \quad {\scriptscriptstyle [0.99,1.01] }$ & $0.97 \quad {\scriptscriptstyle [0.97,1.03] }$ & $1.2 \quad {\scriptscriptstyle [0.8,1.2] }$ & $1.1 \quad {\scriptscriptstyle [0.8,1.2] }$ & $\mathbf{ 2.0 } \quad {\scriptscriptstyle [0.2,1.7] }$ & $0.0 \quad {\scriptscriptstyle [0.0,4.4] }$ \cr       
-6000 & $1.0 \quad {\scriptscriptstyle [0.99,1.01] }$ & $\mathbf{ 1.04 } \quad {\scriptscriptstyle [0.97,1.03] }$ & $1.0 \quad {\scriptscriptstyle [0.8,1.2] }$ & $0.9 \quad {\scriptscriptstyle [0.8,1.2] }$ & $0.7 \quad {\scriptscriptstyle [0.2,1.7] }$ & $0.0 \quad {\scriptscriptstyle [0.0,4.4] }$ \cr       
-4800 & $0.99 \quad {\scriptscriptstyle [0.99,1.01] }$ & $0.98 \quad {\scriptscriptstyle [0.97,1.03] }$ & $1.1 \quad {\scriptscriptstyle [0.8,1.2] }$ & $0.9 \quad {\scriptscriptstyle [0.7,1.2] }$ & $0.7 \quad {\scriptscriptstyle [0.2,1.7] }$ & $0.0 \quad {\scriptscriptstyle [0.0,4.3] }$ \cr       
-3600 & $\mathbf{ 0.97 } \quad {\scriptscriptstyle [0.99,1.01] }$ & $\mathbf{ 0.96 } \quad {\scriptscriptstyle [0.97,1.03] }$ & $1.0 \quad {\scriptscriptstyle [0.8,1.2] }$ & $0.9 \quad {\scriptscriptstyle [0.7,1.2] }$ & $0.6 \quad {\scriptscriptstyle [0.2,1.7] }$ & $2.2 \quad {\scriptscriptstyle [0.0,4.3] }$ \cr       
-2400 & $\mathbf{ 0.97 } \quad {\scriptscriptstyle [0.99,1.01] }$ & $\mathbf{ 0.94 } \quad {\scriptscriptstyle [0.97,1.03] }$ & $\mathbf{ 0.8 } \quad {\scriptscriptstyle [0.8,1.2] }$ & $0.9 \quad {\scriptscriptstyle [0.8,1.2] }$ & $\mathbf{ 0.2 } \quad {\scriptscriptstyle [0.2,1.9] }$ & $0.0 \quad {\scriptscriptstyle [0.0,4.2] }$ \cr       
-1200 & $\mathbf{ 0.96 } \quad {\scriptscriptstyle [0.99,1.01] }$ & $\mathbf{ 0.94 } \quad {\scriptscriptstyle [0.97,1.03] }$ & $1.0 \quad {\scriptscriptstyle [0.8,1.2] }$ & $1.0 \quad {\scriptscriptstyle [0.8,1.2] }$ & $0.8 \quad {\scriptscriptstyle [0.2,1.9] }$ & $2.1 \quad {\scriptscriptstyle [0.0,4.2] }$ \cr       
0 & $\mathbf{ 0.98 } \quad {\scriptscriptstyle [0.99,1.01] }$ & $\mathbf{ 0.96 } \quad {\scriptscriptstyle [0.97,1.03] }$ & $0.9 \quad {\scriptscriptstyle [0.8,1.2] }$ & $0.8 \quad {\scriptscriptstyle [0.8,1.2] }$ & $0.4 \quad {\scriptscriptstyle [0.4,1.9] }$ & $2.1 \quad {\scriptscriptstyle [0.0,4.2] }$ \cr       
1200 & $0.99 \quad {\scriptscriptstyle [0.99,1.01] }$ & $\mathbf{ 0.96 } \quad {\scriptscriptstyle [0.97,1.03] }$ & $1.2 \quad {\scriptscriptstyle [0.8,1.2] }$ & $\mathbf{ 1.3 } \quad {\scriptscriptstyle [0.8,1.2] }$ & $1.1 \quad {\scriptscriptstyle [0.2,1.7] }$ & $0.0 \quad {\scriptscriptstyle [0.0,4.2] }$ \cr       
2400 & $\mathbf{ 0.99 } \quad {\scriptscriptstyle [0.99,1.01] }$ & $\mathbf{ 0.96 } \quad {\scriptscriptstyle [0.97,1.03] }$ & $0.9 \quad {\scriptscriptstyle [0.8,1.2] }$ & $0.9 \quad {\scriptscriptstyle [0.8,1.2] }$ & $0.4 \quad {\scriptscriptstyle [0.2,1.7] }$ & $0.0 \quad {\scriptscriptstyle [0.0,4.3] }$ \cr       
3600 & $0.99 \quad {\scriptscriptstyle [0.99,1.01] }$ & $1.01 \quad {\scriptscriptstyle [0.97,1.03] }$ & $0.9 \quad {\scriptscriptstyle [0.8,1.2] }$ & $0.9 \quad {\scriptscriptstyle [0.7,1.2] }$ & $1.5 \quad {\scriptscriptstyle [0.2,1.7] }$ & $0.0 \quad {\scriptscriptstyle [0.0,4.3] }$ \cr       
4800 & $1.01 \quad {\scriptscriptstyle [0.99,1.01] }$ & $\mathbf{ 1.04 } \quad {\scriptscriptstyle [0.97,1.03] }$ & $0.9 \quad {\scriptscriptstyle [0.8,1.2] }$ & $0.9 \quad {\scriptscriptstyle [0.8,1.2] }$ & $\mathbf{ 0.2 } \quad {\scriptscriptstyle [0.2,1.7] }$ & $0.0 \quad {\scriptscriptstyle [0.0,4.4] }$ \cr       
6000 & $1.0 \quad {\scriptscriptstyle [0.99,1.01] }$ & $1.0 \quad {\scriptscriptstyle [0.97,1.03] }$ & $1.2 \quad {\scriptscriptstyle [0.8,1.2] }$ & $1.0 \quad {\scriptscriptstyle [0.8,1.2] }$ & $0.4 \quad {\scriptscriptstyle [0.2,1.8] }$ & $0.0 \quad {\scriptscriptstyle [0.0,4.4] }$ \cr       
7200 & $1.01 \quad {\scriptscriptstyle [0.99,1.01] }$ & $1.0 \quad {\scriptscriptstyle [0.97,1.03] }$ & $1.2 \quad {\scriptscriptstyle [0.8,1.2] }$ & $1.1 \quad {\scriptscriptstyle [0.7,1.2] }$ & $1.1 \quad {\scriptscriptstyle [0.2,1.8] }$ & $\mathbf{ 4.4 } \quad {\scriptscriptstyle [0.0,4.4] }$ \cr       
8400 & $1.01 \quad {\scriptscriptstyle [0.99,1.01] }$ & $1.02 \quad {\scriptscriptstyle [0.97,1.03] }$ & $1.0 \quad {\scriptscriptstyle [0.8,1.2] }$ & $0.9 \quad {\scriptscriptstyle [0.8,1.2] }$ & $0.7 \quad {\scriptscriptstyle [0.2,1.8] }$ & $2.2 \quad {\scriptscriptstyle [0.0,4.4] }$ \cr       
9600 & $1.0 \quad {\scriptscriptstyle [0.99,1.01] }$ & $1.03 \quad {\scriptscriptstyle [0.97,1.03] }$ & $1.0 \quad {\scriptscriptstyle [0.8,1.2] }$ & $1.0 \quad {\scriptscriptstyle [0.8,1.2] }$ & $1.1 \quad {\scriptscriptstyle [0.2,1.8] }$ & $0.0 \quad {\scriptscriptstyle [0.0,4.4] }$ \cr       
10800 & $1.0 \quad {\scriptscriptstyle [0.99,1.01] }$ & $1.0 \quad {\scriptscriptstyle [0.97,1.03] }$ & $\mathbf{ 0.8 } \quad {\scriptscriptstyle [0.8,1.2] }$ & $\mathbf{ 0.7 } \quad {\scriptscriptstyle [0.8,1.2] }$ & $1.8 \quad {\scriptscriptstyle [0.2,1.8] }$ & $2.2 \quad {\scriptscriptstyle [0.0,4.4] }$ \cr       
12000 & $1.01 \quad {\scriptscriptstyle [0.99,1.01] }$ & $\mathbf{ 1.05 } \quad {\scriptscriptstyle [0.97,1.03] }$ & $1.1 \quad {\scriptscriptstyle [0.8,1.2] }$ & $1.2 \quad {\scriptscriptstyle [0.7,1.2] }$ & $1.1 \quad {\scriptscriptstyle [0.2,1.7] }$ & $0.0 \quad {\scriptscriptstyle [0.0,4.3] }$ \cr       
13200 & $1.01 \quad {\scriptscriptstyle [0.99,1.01] }$ & $1.02 \quad {\scriptscriptstyle [0.97,1.03] }$ & $1.1 \quad {\scriptscriptstyle [0.8,1.2] }$ & $1.2 \quad {\scriptscriptstyle [0.8,1.2] }$ & $0.9 \quad {\scriptscriptstyle [0.2,1.7] }$ & $0.0 \quad {\scriptscriptstyle [0.0,4.3] }$ \cr       
14400 & $1.01 \quad {\scriptscriptstyle [0.99,1.01] }$ & $0.99 \quad {\scriptscriptstyle [0.97,1.03] }$ & $1.1 \quad {\scriptscriptstyle [0.8,1.2] }$ & $1.2 \quad {\scriptscriptstyle [0.8,1.2] }$ & $\mathbf{ 1.9 } \quad {\scriptscriptstyle [0.2,1.7] }$ & $2.1 \quad {\scriptscriptstyle [0.0,4.3] }$ \cr       
15600 & $1.0 \quad {\scriptscriptstyle [0.99,1.01] }$ & $1.01 \quad {\scriptscriptstyle [0.97,1.03] }$ & $1.0 \quad {\scriptscriptstyle [0.8,1.2] }$ & $0.9 \quad {\scriptscriptstyle [0.8,1.2] }$ & $1.1 \quad {\scriptscriptstyle [0.2,1.9] }$ & $0.0 \quad {\scriptscriptstyle [0.0,4.2] }$ \cr       
16800 & $1.0 \quad {\scriptscriptstyle [0.99,1.01] }$ & $0.97 \quad {\scriptscriptstyle [0.97,1.03] }$ & $0.9 \quad {\scriptscriptstyle [0.8,1.2] }$ & $0.8 \quad {\scriptscriptstyle [0.8,1.2] }$ & $1.3 \quad {\scriptscriptstyle [0.2,1.7] }$ & $0.0 \quad {\scriptscriptstyle [0.0,4.3] }$ \cr

\mr
$\sigma{\scriptstyle\left[ \text{FAR}^{\text{slag}}_{\text{emp}} \right] }$ &
$ {\scriptstyle \pm 0.01(6) } $ &
$ {\scriptstyle \pm 0.03(1) } $ &
$ {\scriptstyle \pm 0.1(2) } $ &
$ {\scriptstyle \pm 0.1(6) } $ &
$ {\scriptstyle \pm 0.4(6) } $ &
$ {\scriptstyle \pm 1.(2) } $ \cr

$\sigma{\scriptstyle\left[ \text{FAR}^{\text{slag}}_{\text{th}} \right] }$ &
$ {\scriptstyle \pm 0.007 } $ &
$ {\scriptstyle \pm 0.02(0) } $ &
$ {\scriptstyle \pm 0.1(0) } $ &
$ {\scriptstyle \pm 0.1(5) } $ &
$ {\scriptstyle \pm 0.4(6) } $ &
$ {\scriptstyle \pm 1.(4) } $ \cr

\mr
$\text{FAR}^{\text{bkg}}_{\text{emp}} \pm \sigma^{\text{bkg}}_{\text{emp}}$ &
$ {\scriptstyle (1.001 \pm 0.001) } $ &
$ {\scriptstyle (1.000 \pm 0.004) } $ &
$ {\scriptstyle (1.01  \pm  0.02) } $ &
$ {\scriptstyle (0.98  \pm  0.03) } $ &
$ {\scriptstyle (0.99  \pm  0.09) } $ &
$ {\scriptstyle (0.9   \pm   0.3) } $ \cr

$\text{FAR}^{\text{bkg}}_{\text{th}} \pm \sigma^{\text{bkg}}_{\text{th}}$ &
$ {\scriptstyle (1.00144 \pm 0.00001) } $ &
$ {\scriptstyle (1.00041 \pm 0.00004) } $ &
$ {\scriptstyle (1.0083  \pm  0.0002) } $ &
$ {\scriptstyle (0.9835  \pm  0.0003) } $ &
$ {\scriptstyle (0.9897  \pm  0.0009) } $ &
$ {\scriptstyle (0.910   \pm   0.003) } $ \cr

\br

\end{tabular}
\caption{Comparison of FAR values at fixed rho thredholds as a function of the average shift in time of disjoint sets of time slides. The time shift is applied to the Hanford data stream. The FAR estimates that fluctuate more than 2 standard deviations from the mean FAR (last row) are highlighted in bold.}
\label{tab:FARslagsLH}
\end{table}

\begin{table}
\footnotesize
\lineup
\renewcommand{\arraystretch}{1.2}
\begin{tabular}{@{}*{6}{c}}

\br                              
 & $\text{B}_{\text{ BBH}}$  & $\text{C}_{\text{ BBH}}$ &  $\text{D}_{\text{ BBH}}$ & $\text{E}_{\text{ BBH}}$ & $\text{F}_{\text{ BBH}}$ \cr

\mr
\multirow{2}{*}{Time shift [s]} &
\multicolumn{5}{c}{$ \text{FAR}^{\text{slag}}_{\text{emp}} \quad {\scriptscriptstyle [\text{FAR}_{\text{slag, th}}^{5\%}, \, \text{FAR}_{\text{slag, th}}^{95\%}] } $} \\
 & {\tiny [1/week]} & {\tiny [1/6 month]} & {\tiny [1/year]} & {\tiny [1/10years]} & {\tiny [1/100years]} \\

\mr

-16800 & $0.997 \quad {\scriptscriptstyle [0.966,1.035] }$ & $\mathbf{ 1.211 } \quad {\scriptscriptstyle [0.827,1.187] }$ & $1.22 \quad {\scriptscriptstyle [0.73,1.24] }$ & $1.4 \quad {\scriptscriptstyle [0.23,1.86] }$ & $0.0 \quad {\scriptscriptstyle [0.0,4.7] }$ \cr       
-15600 & $0.984 \quad {\scriptscriptstyle [0.966,1.034] }$ & $0.857 \quad {\scriptscriptstyle [0.822,1.181] }$ & $1.01 \quad {\scriptscriptstyle [0.75,1.23] }$ & $1.39 \quad {\scriptscriptstyle [0.23,1.85] }$ & $2.3 \quad {\scriptscriptstyle [0.0,4.6] }$ \cr       
-14400 & $1.013 \quad {\scriptscriptstyle [0.966,1.035] }$ & $1.086 \quad {\scriptscriptstyle [0.823,1.177] }$ & $1.06 \quad {\scriptscriptstyle [0.74,1.24] }$ & $0.46 \quad {\scriptscriptstyle [0.23,1.83] }$ & $0.0 \quad {\scriptscriptstyle [0.0,4.6] }$ \cr       
-13200 & $1.003 \quad {\scriptscriptstyle [0.966,1.034] }$ & $0.879 \quad {\scriptscriptstyle [0.834,1.172] }$ & $0.8 \quad {\scriptscriptstyle [0.76,1.24] }$ & $1.13 \quad {\scriptscriptstyle [0.23,1.8] }$ & $2.3 \quad {\scriptscriptstyle [0.0,4.5] }$ \cr       
-12000 & $0.994 \quad {\scriptscriptstyle [0.967,1.035] }$ & $\mathbf{ 1.197 } \quad {\scriptscriptstyle [0.827,1.174] }$ & $1.19 \quad {\scriptscriptstyle [0.75,1.24] }$ & $0.67 \quad {\scriptscriptstyle [0.22,1.79] }$ & $0.0 \quad {\scriptscriptstyle [0.0,4.5] }$ \cr       
-10800 & $0.998 \quad {\scriptscriptstyle [0.966,1.034] }$ & $1.038 \quad {\scriptscriptstyle [0.826,1.183] }$ & $1.06 \quad {\scriptscriptstyle [0.75,1.23] }$ & $0.45 \quad {\scriptscriptstyle [0.22,1.79] }$ & $0.0 \quad {\scriptscriptstyle [0.0,4.5] }$ \cr       
-9600 & $1.009 \quad {\scriptscriptstyle [0.966,1.035] }$ & $1.001 \quad {\scriptscriptstyle [0.834,1.179] }$ & $1.08 \quad {\scriptscriptstyle [0.75,1.23] }$ & $1.34 \quad {\scriptscriptstyle [0.22,1.78] }$ & $0.0 \quad {\scriptscriptstyle [0.0,4.5] }$ \cr       
-8400 & $1.024 \quad {\scriptscriptstyle [0.966,1.034] }$ & $0.981 \quad {\scriptscriptstyle [0.826,1.179] }$ & $0.85 \quad {\scriptscriptstyle [0.76,1.24] }$ & $0.88 \quad {\scriptscriptstyle [0.22,1.76] }$ & $0.0 \quad {\scriptscriptstyle [0.0,4.4] }$ \cr       
-7200 & $0.999 \quad {\scriptscriptstyle [0.967,1.034] }$ & $\mathbf{ 0.819 } \quad {\scriptscriptstyle [0.83,1.179] }$ & $0.86 \quad {\scriptscriptstyle [0.75,1.23] }$ & $1.09 \quad {\scriptscriptstyle [0.22,1.75] }$ & $2.2 \quad {\scriptscriptstyle [0.0,4.4] }$ \cr       
-6000 & $1.016 \quad {\scriptscriptstyle [0.966,1.034] }$ & $\mathbf{ 1.242 } \quad {\scriptscriptstyle [0.839,1.177] }$ & $1.03 \quad {\scriptscriptstyle [0.75,1.23] }$ & $0.44 \quad {\scriptscriptstyle [0.22,1.74] }$ & $0.0 \quad {\scriptscriptstyle [0.0,4.4] }$ \cr       
-4800 & $1.02 \quad {\scriptscriptstyle [0.967,1.034] }$ & $0.933 \quad {\scriptscriptstyle [0.835,1.172] }$ & $0.96 \quad {\scriptscriptstyle [0.75,1.24] }$ & $1.52 \quad {\scriptscriptstyle [0.22,1.74] }$ & $0.0 \quad {\scriptscriptstyle [0.0,4.3] }$ \cr       
-3600 & $0.987 \quad {\scriptscriptstyle [0.967,1.033] }$ & $1.025 \quad {\scriptscriptstyle [0.831,1.176] }$ & $1.0 \quad {\scriptscriptstyle [0.77,1.23] }$ & $1.29 \quad {\scriptscriptstyle [0.42,1.73] }$ & $4.3 \quad {\scriptscriptstyle [0.0,4.3] }$ \cr       
-2400 & $1.019 \quad {\scriptscriptstyle [0.968,1.034] }$ & $0.91 \quad {\scriptscriptstyle [0.836,1.174] }$ & $0.83 \quad {\scriptscriptstyle [0.75,1.23] }$ & $0.85 \quad {\scriptscriptstyle [0.42,1.69] }$ & $0.0 \quad {\scriptscriptstyle [0.0,4.2] }$ \cr       
-1200 & $\mathbf{ 0.939 } \quad {\scriptscriptstyle [0.968,1.033] }$ & $\mathbf{ 0.836 } \quad {\scriptscriptstyle [0.836,1.17] }$ & $0.89 \quad {\scriptscriptstyle [0.76,1.24] }$ & $1.04 \quad {\scriptscriptstyle [0.42,1.88] }$ & $0.0 \quad {\scriptscriptstyle [0.0,4.2] }$ \cr       
0 & $0.986 \quad {\scriptscriptstyle [0.967,1.033] }$ & $\mathbf{ 0.831 } \quad {\scriptscriptstyle [0.831,1.173] }$ & $0.82 \quad {\scriptscriptstyle [0.76,1.23] }$ & $0.83 \quad {\scriptscriptstyle [0.21,1.87] }$ & $0.0 \quad {\scriptscriptstyle [0.0,4.2] }$ \cr       
1200 & $1.006 \quad {\scriptscriptstyle [0.967,1.033] }$ & $1.027 \quad {\scriptscriptstyle [0.836,1.175] }$ & $1.02 \quad {\scriptscriptstyle [0.75,1.23] }$ & $1.48 \quad {\scriptscriptstyle [0.21,1.91] }$ & $2.1 \quad {\scriptscriptstyle [0.0,4.2] }$ \cr       
2400 & $1.012 \quad {\scriptscriptstyle [0.967,1.033] }$ & $1.038 \quad {\scriptscriptstyle [0.835,1.177] }$ & $1.14 \quad {\scriptscriptstyle [0.76,1.22] }$ & $1.28 \quad {\scriptscriptstyle [0.21,1.71] }$ & $2.1 \quad {\scriptscriptstyle [0.0,4.3] }$ \cr       
3600 & $1.01 \quad {\scriptscriptstyle [0.967,1.034] }$ & $1.09 \quad {\scriptscriptstyle [0.831,1.176] }$ & $1.0 \quad {\scriptscriptstyle [0.77,1.23] }$ & $0.65 \quad {\scriptscriptstyle [0.22,1.73] }$ & $0.0 \quad {\scriptscriptstyle [0.0,4.3] }$ \cr       
4800 & $1.002 \quad {\scriptscriptstyle [0.967,1.035] }$ & $1.024 \quad {\scriptscriptstyle [0.839,1.177] }$ & $1.12 \quad {\scriptscriptstyle [0.75,1.23] }$ & $0.65 \quad {\scriptscriptstyle [0.22,1.74] }$ & $0.0 \quad {\scriptscriptstyle [0.0,4.4] }$ \cr       
6000 & $0.969 \quad {\scriptscriptstyle [0.967,1.034] }$ & $0.844 \quad {\scriptscriptstyle [0.833,1.173] }$ & $0.8 \quad {\scriptscriptstyle [0.76,1.23] }$ & $1.1 \quad {\scriptscriptstyle [0.22,1.75] }$ & $0.0 \quad {\scriptscriptstyle [0.0,4.4] }$ \cr       
7200 & $0.994 \quad {\scriptscriptstyle [0.966,1.034] }$ & $0.991 \quad {\scriptscriptstyle [0.826,1.178] }$ & $0.87 \quad {\scriptscriptstyle [0.76,1.24] }$ & $0.66 \quad {\scriptscriptstyle [0.22,1.76] }$ & $0.0 \quad {\scriptscriptstyle [0.0,4.4] }$ \cr       
8400 & $0.995 \quad {\scriptscriptstyle [0.965,1.034] }$ & $1.067 \quad {\scriptscriptstyle [0.836,1.177] }$ & $0.76 \quad {\scriptscriptstyle [0.76,1.24] }$ & $0.88 \quad {\scriptscriptstyle [0.22,1.76] }$ & $0.0 \quad {\scriptscriptstyle [0.0,4.4] }$ \cr       
9600 & $1.007 \quad {\scriptscriptstyle [0.966,1.034] }$ & $0.946 \quad {\scriptscriptstyle [0.836,1.177] }$ & $0.93 \quad {\scriptscriptstyle [0.74,1.24] }$ & $1.76 \quad {\scriptscriptstyle [0.22,1.76] }$ & $\mathbf{ 6.6 } \quad {\scriptscriptstyle [0.0,4.4] }$ \cr       
10800 & $0.972 \quad {\scriptscriptstyle [0.966,1.034] }$ & $0.909 \quad {\scriptscriptstyle [0.833,1.172] }$ & $1.04 \quad {\scriptscriptstyle [0.76,1.23] }$ & $0.66 \quad {\scriptscriptstyle [0.22,1.75] }$ & $0.0 \quad {\scriptscriptstyle [0.0,4.4] }$ \cr       
12000 & $0.984 \quad {\scriptscriptstyle [0.967,1.034] }$ & $0.964 \quad {\scriptscriptstyle [0.834,1.181] }$ & $0.9 \quad {\scriptscriptstyle [0.75,1.24] }$ & $0.43 \quad {\scriptscriptstyle [0.22,1.73] }$ & $0.0 \quad {\scriptscriptstyle [0.0,4.3] }$ \cr       
13200 & $1.019 \quad {\scriptscriptstyle [0.968,1.034] }$ & $\mathbf{ 1.189 } \quad {\scriptscriptstyle [0.835,1.178] }$ & $0.99 \quad {\scriptscriptstyle [0.76,1.23] }$ & $1.07 \quad {\scriptscriptstyle [0.43,1.93] }$ & $0.0 \quad {\scriptscriptstyle [0.0,4.3] }$ \cr       
14400 & $1.011 \quad {\scriptscriptstyle [0.967,1.033] }$ & $1.025 \quad {\scriptscriptstyle [0.833,1.174] }$ & $1.01 \quad {\scriptscriptstyle [0.76,1.22] }$ & $1.71 \quad {\scriptscriptstyle [0.21,1.92] }$ & $0.0 \quad {\scriptscriptstyle [0.0,4.3] }$ \cr       
15600 & $1.011 \quad {\scriptscriptstyle [0.968,1.033] }$ & $0.965 \quad {\scriptscriptstyle [0.828,1.178] }$ & $1.13 \quad {\scriptscriptstyle [0.75,1.21] }$ & $0.85 \quad {\scriptscriptstyle [0.21,1.91] }$ & $2.1 \quad {\scriptscriptstyle [0.0,4.2] }$ \cr       
16800 & $1.022 \quad {\scriptscriptstyle [0.967,1.033] }$ & $1.098 \quad {\scriptscriptstyle [0.831,1.172] }$ & $\mathbf{ 1.24 } \quad {\scriptscriptstyle [0.76,1.22] }$ & $1.07 \quad {\scriptscriptstyle [0.21,1.72] }$ & $2.1 \quad {\scriptscriptstyle [0.0,4.3] }$ \cr

\mr
$\sigma{\scriptstyle\left[ \text{FAR}^{\text{slag}}_{\text{emp}} \right] }$ &
$ {\scriptstyle \pm 0.01(8)} $ &
$ {\scriptstyle \pm 0.1(2)}  $ &
$ {\scriptstyle \pm 0.1(3)}  $ &
$ {\scriptstyle \pm 0.3(8)}  $ &
$ {\scriptstyle \pm 1.(6)}   $ \cr

$\sigma{\scriptstyle\left[ \text{FAR}^{\text{slag}}_{\text{th}} \right] }$ &
$ {\scriptstyle \pm 0.02} $ &
$ {\scriptstyle \pm 0.1(0)} $ &
$ {\scriptstyle \pm 0.1(5)} $ &
$ {\scriptstyle \pm 0.4(7)} $ &
$ {\scriptstyle \pm 1.(4)} $ \cr

\mr
$\text{FAR}^{\text{bkg}}_{\text{emp}} \pm \sigma^{\text{bkg}}_{\text{emp}}$ &
$ {\scriptstyle (1.000 \pm 0.004) } $ &
$ {\scriptstyle (1.00  \pm 0.02) }  $ &
$ {\scriptstyle (0.99  \pm 0.03) }  $ &
$ {\scriptstyle (1.00  \pm 0.09) }  $ &
$ {\scriptstyle (0.9   \pm 0.3) }   $ \cr

$\text{FAR}^{\text{bkg}}_{\text{th}} \pm \sigma^{\text{bkg}}_{\text{th}}$ &
$ {\scriptstyle (1.00007 \pm 0.00004) } $ &
$ {\scriptstyle (1.0003  \pm 0.0002) }  $ &
$ {\scriptstyle (0.9862  \pm 0.0003) }  $ &
$ {\scriptstyle (1.0008  \pm 0.0009) }  $ &
$ {\scriptstyle (0.898   \pm 0.003) }   $ \cr

\br

\end{tabular}
\caption{Comparison of FAR values at fixed rho thredholds as a function of the average shift in time of disjoint sets of time slides. The time shift is applied to the Hanford data stream. The FAR estimates that fluctuate more than 2 standard deviations from the mean FAR (last row) are highlighted in bold.}
\label{tab:FARslagsBBH}
\end{table}

\begin{table}
\footnotesize
\lineup
\renewcommand{\arraystretch}{1.2}
\begin{tabular}{@{}*{7}{c}}

\br                              
 & $\text{A}_{\text{ LHV}}$ & $\text{B}_{\text{ LHV}}$ & $\text{C}_{\text{ LHV}}$ & $\text{D}_{\text{ LHV}}$ & $\text{E}_{\text{ LHV}}$ & $\text{F}_{\text{ LHV}}$ \cr

\mr
\multirow{2}{*}{Time shift [s]} &
\multicolumn{6}{c}{$ \text{FAR}^{\text{slag}}_{\text{emp}} \quad {\scriptscriptstyle [\text{FAR}_{\text{slag, th}}^{5\%}, \, \text{FAR}_{\text{slag, th}}^{95\%}] } $} \\
 & {\tiny [1/day]} & {\tiny [1/week]} & {\tiny [1/6 month]} & {\tiny [1/year]} & {\tiny [1/10years]} & {\tiny [1/100years]} \\

\mr
0 & $0.996 \quad {\scriptscriptstyle [0.987,1.012] }$ & $0.98 \quad {\scriptscriptstyle [0.97,1.03] }$ & $0.99 \quad {\scriptscriptstyle [0.84,1.18] }$ & $0.92 \quad {\scriptscriptstyle [0.76,1.24] }$ & $1.26 \quad {\scriptscriptstyle [0.42,1.89] }$ & $2.1 \quad {\scriptscriptstyle [0.0,4.2] }$ \cr       
1200 & $\mathbf{ 0.977 } \quad {\scriptscriptstyle [0.987,1.012] }$ & $0.97 \quad {\scriptscriptstyle [0.97,1.03] }$ & $1.05 \quad {\scriptscriptstyle [0.84,1.17] }$ & $1.08 \quad {\scriptscriptstyle [0.76,1.26] }$ & $\mathbf{ 0.22 } \quad {\scriptscriptstyle [0.22,1.74] }$ & $0.0 \quad {\scriptscriptstyle [0.0,4.3] }$ \cr       
3600 & $\mathbf{ 0.967 } \quad {\scriptscriptstyle [0.987,1.012] }$ & $\mathbf{ 0.95 } \quad {\scriptscriptstyle [0.97,1.03] }$ & $0.9 \quad {\scriptscriptstyle [0.84,1.17] }$ & $1.05 \quad {\scriptscriptstyle [0.75,1.25] }$ & $1.72 \quad {\scriptscriptstyle [0.22,1.72] }$ & $0.0 \quad {\scriptscriptstyle [0.0,4.3] }$ \cr       
6000 & $0.99 \quad {\scriptscriptstyle [0.987,1.013] }$ & $0.99 \quad {\scriptscriptstyle [0.97,1.03] }$ & $0.94 \quad {\scriptscriptstyle [0.84,1.17] }$ & $0.89 \quad {\scriptscriptstyle [0.76,1.24] }$ & $0.65 \quad {\scriptscriptstyle [0.22,1.74] }$ & $0.0 \quad {\scriptscriptstyle [0.0,4.3] }$ \cr       
7200 & $\mathbf{ 1.018 } \quad {\scriptscriptstyle [0.987,1.012] }$ & $\mathbf{ 1.05 } \quad {\scriptscriptstyle [0.97,1.03] }$ & $1.04 \quad {\scriptscriptstyle [0.83,1.17] }$ & $0.96 \quad {\scriptscriptstyle [0.77,1.25] }$ & $1.53 \quad {\scriptscriptstyle [0.22,1.75] }$ & $2.2 \quad {\scriptscriptstyle [0.0,4.4] }$ \cr       
8400 & $0.996 \quad {\scriptscriptstyle [0.987,1.012] }$ & $0.99 \quad {\scriptscriptstyle [0.97,1.03] }$ & $0.95 \quad {\scriptscriptstyle [0.83,1.17] }$ & $0.95 \quad {\scriptscriptstyle [0.76,1.24] }$ & $1.08 \quad {\scriptscriptstyle [0.22,1.73] }$ & $0.0 \quad {\scriptscriptstyle [0.0,4.3] }$ \cr       
9600 & $1.008 \quad {\scriptscriptstyle [0.987,1.012] }$ & $1.02 \quad {\scriptscriptstyle [0.97,1.03] }$ & $0.97 \quad {\scriptscriptstyle [0.83,1.18] }$ & $1.01 \quad {\scriptscriptstyle [0.76,1.25] }$ & $0.43 \quad {\scriptscriptstyle [0.22,1.73] }$ & $0.0 \quad {\scriptscriptstyle [0.0,4.3] }$ \cr       
10800 & $\mathbf{ 1.026 } \quad {\scriptscriptstyle [0.987,1.012] }$ & $1.03 \quad {\scriptscriptstyle [0.97,1.03] }$ & $1.07 \quad {\scriptscriptstyle [0.83,1.18] }$ & $1.11 \quad {\scriptscriptstyle [0.77,1.24] }$ & $1.07 \quad {\scriptscriptstyle [0.21,1.71] }$ & $4.3 \quad {\scriptscriptstyle [0.0,4.3] }$ \cr       
12000 & $\mathbf{ 1.015 } \quad {\scriptscriptstyle [0.987,1.013] }$ & $1.03 \quad {\scriptscriptstyle [0.97,1.03] }$ & $0.91 \quad {\scriptscriptstyle [0.83,1.17] }$ & $0.9 \quad {\scriptscriptstyle [0.77,1.24] }$ & $0.64 \quad {\scriptscriptstyle [0.21,1.71] }$ & $2.1 \quad {\scriptscriptstyle [0.0,4.3] }$ \cr       
13200 & $\mathbf{ 1.016 } \quad {\scriptscriptstyle [0.987,1.012] }$ & $1.01 \quad {\scriptscriptstyle [0.97,1.03] }$ & $1.15 \quad {\scriptscriptstyle [0.83,1.17] }$ & $1.02 \quad {\scriptscriptstyle [0.77,1.26] }$ & $0.64 \quad {\scriptscriptstyle [0.21,1.92] }$ & $0.0 \quad {\scriptscriptstyle [0.0,4.3] }$ \cr       
14400 & $0.993 \quad {\scriptscriptstyle [0.987,1.012] }$ & $0.98 \quad {\scriptscriptstyle [0.97,1.03] }$ & $0.98 \quad {\scriptscriptstyle [0.84,1.17] }$ & $0.97 \quad {\scriptscriptstyle [0.76,1.25] }$ & $1.27 \quad {\scriptscriptstyle [0.21,1.9] }$ & $0.0 \quad {\scriptscriptstyle [0.0,4.2] }$ \cr       
16800 & $0.996 \quad {\scriptscriptstyle [0.987,1.012] }$ & $0.99 \quad {\scriptscriptstyle [0.97,1.03] }$ & $1.03 \quad {\scriptscriptstyle [0.83,1.18] }$ & $1.11 \quad {\scriptscriptstyle [0.77,1.25] }$ & $1.46 \quad {\scriptscriptstyle [0.21,1.88] }$ & $0.0 \quad {\scriptscriptstyle [0.0,4.2] }$ \cr

\mr
$\sigma{\scriptstyle\left[ \text{FAR}^{\text{slag}}_{\text{emp}} \right] }$ &
$ {\scriptstyle \pm 0.01(7)} $ &
$ {\scriptstyle \pm 0.02(7)} $ &
$ {\scriptstyle \pm 0.06(9)} $ &
$ {\scriptstyle \pm 0.07(5)} $ &
$ {\scriptstyle \pm 0.4(6)}  $ &
$ {\scriptstyle \pm 1.(4)}  $ \cr

$\sigma{\scriptstyle\left[ \text{FAR}^{\text{slag}}_{\text{th}} \right] }$ &
$ {\scriptstyle \pm 0.007(7)} $ &
$ {\scriptstyle \pm 0.02(0)}  $ &
$ {\scriptstyle \pm 0.1(0)}   $ &
$ {\scriptstyle \pm 0.1(5)}   $ &
$ {\scriptstyle \pm 0.4(6)}   $ &
$ {\scriptstyle \pm 1.(4)}    $ \cr

\mr
$\text{FAR}^{\text{bkg}}_{\text{emp}} \pm \sigma^{\text{bkg}}_{\text{emp}}$ &
$ {\scriptstyle (1.000 \pm 0.002) } $ &
$ {\scriptstyle (0.999 \pm 0.006) } $ &
$ {\scriptstyle (1.00  \pm 0.03) }  $ &
$ {\scriptstyle (1.00  \pm 0.04) }  $ &
$ {\scriptstyle (1.0   \pm 0.1) }   $ &
$ {\scriptstyle (0.9   \pm 0.4) }   $ \cr

$\text{FAR}^{\text{bkg}}_{\text{th}} \pm \sigma^{\text{bkg}}_{\text{th}}$ &
$ {\scriptstyle (0.99973 \pm 0.00002) } $ &
$ {\scriptstyle (0.99945 \pm 0.00006) } $ &
$ {\scriptstyle (1.0001  \pm 0.0003) }  $ &
$ {\scriptstyle (0.9996  \pm 0.0004) }  $ &
$ {\scriptstyle (0.9996  \pm 0.0004) }  $ &
$ {\scriptstyle (0.895   \pm 0.004) }   $ \cr

\br

\end{tabular}
\caption{Comparison of FAR values at fixed rho thredholds as a function of the average shift in time of disjoint sets of time slides. The time shift is applied to the Hanford data stream. The FAR estimates that fluctuate more than 2 standard deviations from the mean FAR (last row) are highlighted in bold.}
\label{tab:FARslagsLHV}
\end{table}


\newpage

\bibliographystyle{unsrt}
\bibliography{bibliography}

\begin{thebibliography}{10}

\bibitem{aLIGO}
J.~Aasi et~al.
\newblock {Advanced LIGO}.
\newblock {\em Class. Quant. Grav.}, 32:074001, 2015.

\bibitem{aVIRGO}
F.~Acernese et~al.
\newblock {Advanced Virgo: a second-generation interferometric gravitational wave detector}.
\newblock {\em Class. Quant. Grav.}, 32(2):024001, 2015.

\bibitem{GWTC-1}
B.~P. Abbott et~al.
\newblock {GWTC-1: A Gravitational-Wave Transient Catalog of Compact Binary Mergers Observed by LIGO and Virgo during the First and Second Observing Runs}.
\newblock {\em Phys. Rev. X}, 9(3):031040, 2019.

\bibitem{GWTC-2}
R.~Abbott et~al.
\newblock {GWTC-2.1: Deep extended catalog of compact binary coalescences observed by LIGO and Virgo during the first half of the third observing run}.
\newblock {\em Phys. Rev. D}, 109(2):022001, 2024.

\bibitem{GWTC-3}
R.~Abbott et~al.
\newblock {GWTC-3: Compact Binary Coalescences Observed by LIGO and Virgo during the Second Part of the Third Observing Run}.
\newblock {\em Phys. Rev. X}, 13(4):041039, 2023.

\bibitem{GWTC-4results}
The LIGO~Scientific Collaboration, The~Virgo Collaboration, and the KAGRA~Collaboration.
\newblock Gwtc-4.0: Updating the gravitational-wave transient catalog with observations from the first part of the fourth ligo-virgo-kagra observing run, 2025.

\bibitem{GWOSC}
Gravitational wave open science center.

\bibitem{TGR1}
B.~P. Abbott et~al.
\newblock {Tests of General Relativity with the Binary Black Hole Signals from the LIGO-Virgo Catalog GWTC-1}.
\newblock {\em Phys. Rev. D}, 100(10):104036, 2019.

\bibitem{TGR2}
R.~Abbott et~al.
\newblock {Tests of general relativity with binary black holes from the second LIGO-Virgo gravitational-wave transient catalog}.
\newblock {\em Phys. Rev. D}, 103(12):122002, 2021.

\bibitem{TGR3}
R.~Abbott et~al.
\newblock {Tests of General Relativity with GWTC-3}.
\newblock 12 2021.

\bibitem{Abdikamalov2022}
Ernazar Abdikamalov, Giulia Pagliaroli, and David Radice.
\newblock {\em Gravitational Waves from Core-Collapse Supernovae}, pages 909--945.
\newblock Springer Nature Singapore, Singapore, 2022.

\bibitem{GWfromMagnetars2025}
Chris Kouvaris.
\newblock Gravitational waves from magnetars.
\newblock {\em International Journal of Modern Physics D}, 34(09):2550037, 2025.

\bibitem{starquakes2022}
E~Giliberti and G~Cambiotti.
\newblock Starquakes in millisecond pulsars and gravitational waves emission.
\newblock {\em Monthly Notices of the Royal Astronomical Society}, 511(3):3365--3376, 01 2022.

\bibitem{PhysRevD.101.104041}
Michael Ebersold and Shubhanshu Tiwari.
\newblock Search for nonlinear memory from subsolar mass compact binary mergers.
\newblock {\em Phys. Rev. D}, 101:104041, May 2020.

\bibitem{collapsardiskGottlieb2024}
Ore Gottlieb, Amir Levinson, and Yuri Levin.
\newblock In ligo’s sight? vigorous coherent gravitational waves from cooled collapsar disks.
\newblock {\em The Astrophysical Journal Letters}, 972(1):L4, aug 2024.

\bibitem{allsky1}
Benjamin~P. Abbott et~al.
\newblock {All-sky search for short gravitational-wave bursts in the first Advanced LIGO run}.
\newblock {\em Phys. Rev. D}, 95(4):042003, 2017.

\bibitem{allsky2}
B.~P. Abbott et~al.
\newblock {All-Sky Search for Short Gravitational-Wave Bursts in the Second Advanced LIGO and Advanced Virgo Run}.
\newblock {\em Phys. Rev. D}, 100(2):024017, 2019.

\bibitem{allsky3}
R.~Abbott et~al.
\newblock {All-sky search for short gravitational-wave bursts in the third Advanced LIGO and Advanced Virgo run}.
\newblock {\em Phys. Rev. D}, 104(12):122004, 2021.

\bibitem{allsky4}
LIGO~Scientific Collaboration, Virgo Collaboration, and KAGRA Collaboration.
\newblock All-sky search for short gravitational-wave bursts in the first part of the fourth ligo-virgo-kagra observing run, 2025.

\bibitem{long1}
Benjamin~P. Abbott et~al.
\newblock {All-sky search for long-duration gravitational wave transients in the first Advanced LIGO observing run}.
\newblock {\em Class. Quant. Grav.}, 35(6):065009, 2018.

\bibitem{long2}
B.~P. Abbott et~al.
\newblock {All-sky search for long-duration gravitational-wave transients in the second Advanced LIGO observing run}.
\newblock {\em Phys. Rev. D}, 99(10):104033, 2019.

\bibitem{long3}
R.~Abbott et~al.
\newblock {All-sky search for long-duration gravitational-wave bursts in the third Advanced LIGO and Advanced Virgo run}.
\newblock {\em Phys. Rev. D}, 104(10):102001, 2021.

\bibitem{long4}
The LIGO~Scientific Collaboration, the Virgo~Collaboration, and the KAGRA~Collaboration.
\newblock All-sky search for long-duration gravitational-wave transients in the first part of the fourth ligo-virgo-kagra observing run, 2025.

\bibitem{cwb2005}
S.~Klimenko et~al.
\newblock {Constraint likelihood analysis for a network of gravitational wave detectors}.
\newblock {\em Phys. Rev. D}, 72:122002, 2005.

\bibitem{cwb2008}
S~Klimenko et~al.
\newblock A coherent method for detection of gravitational wave bursts.
\newblock {\em Classical and Quantum Gravity}, 25(11):114029, may 2008.

\bibitem{cwb2016}
S.~Klimenko et~al.
\newblock {Method for detection and reconstruction of gravitational wave transients with networks of advanced detectors}.
\newblock {\em Phys. Rev. D}, 93(4):042004, 2016.

\bibitem{gw150914_discovery}
B.~P. Abbott et~al.
\newblock Observation of gravitational waves from a binary black hole merger.
\newblock {\em Phys. Rev. Lett.}, 116:061102, Feb 2016.

\bibitem{gw190521_discovery}
R.~Abbott et~al.
\newblock Gw190521: A binary black hole merger with a total mass of $150 m_{\odot}$.
\newblock {\em Phys. Rev. Lett.}, 125:101102, Sep 2020.

\bibitem{GW231123}
The LIGO~Scientific Collaboration, the Virgo~Collaboration, and the KAGRA~Collaboration.
\newblock Gw231123: a binary black hole merger with total mass 190-265 $m_{\odot}$, 2025.

\bibitem{cwbsftX}
Marco Drago et~al.
\newblock coherent waveburst, a pipeline for unmodeled gravitational-wave data analysis.
\newblock {\em SoftwareX}, 14:100678, 2021.

\bibitem{cwb2023}
Marek~J. Szczepa\ifmmode~\acute{n}\else \'{n}\fi{}czyk, Francesco Salemi, Sophie Bini, Tanmaya Mishra, Gabriele Vedovato, V.~Gayathri, Imre Bartos, Shubhagata Bhaumik, Marco Drago, Odysse Halim, Claudia Lazzaro, Andrea Miani, Edoardo Milotti, Giovanni~A. Prodi, Shubhanshu Tiwari, and Sergey Klimenko.
\newblock Search for gravitational-wave bursts in the third advanced ligo-virgo run with coherent waveburst enhanced by machine learning.
\newblock {\em Phys. Rev. D}, 107:062002, Mar 2023.

\bibitem{cwbxp2022}
Sergey Klimenko.
\newblock Wavescan: multiresolution regression of gravitational-wave data, 2022.

\bibitem{cwbxp2025}
T.~Mishra, S.~Bhaumik, V.~Gayathri, Marek~J. Szczepa\ifmmode~\acute{n}\else \'{n}\fi{}czyk, I.~Bartos, and S.~Klimenko.
\newblock Gravitational waves detected by a burst search in ligo/virgo's third observing run.
\newblock {\em Phys. Rev. D}, 111:023054, Jan 2025.

\bibitem{cwbGMM2022}
Dixeena Lopez, V.~Gayathri, Archana Pai, Ik~Siong Heng, Chris Messenger, and Sagar~Kumar Gupta.
\newblock Utilizing gaussian mixture models in all-sky searches for short-duration gravitational wave bursts.
\newblock {\em Phys. Rev. D}, 105:063024, Mar 2022.

\bibitem{cwbGMM2024}
Leigh Smith, Sayantan Ghosh, Jiyoon Sun, V.~Gayathri, Ik~Siong Heng, and Archana Pai.
\newblock Enhancing search pipelines for short gravitational-wave transients with gaussian mixture modeling.
\newblock {\em Phys. Rev. D}, 110:083032, Oct 2024.

\bibitem{cwbGMM2025}
Sayantan Ghosh, Leigh Smith, Jiyoon Sun, Archana Pai, Ik~Siong Heng, and V.~Gayathri.
\newblock Leveraging cross-detector parameter consistency measures to enhance sensitivities of gravitational-wave searches.
\newblock {\em Phys. Rev. D}, 112:063026, Sep 2025.

\bibitem{cWB-6.4.6}
Sergey Klimenko, Gabriele Vedovato, Valentin Necula, Francesco Salemi, Marco Drago, Rhys Poulton, Sophie Bini, Eric Chassande-Mottin, Claudia Lazzaro, Andrea Miani, Tanmaya Mishra, Brendan O'Brian, Marek Szczepanczyk, Shubhanshu Tiwari, Vaibhav Tiwari, and V.~Gayathri.
\newblock cwb pipeline library: 6.4.6, February 2024.

\bibitem{Marek:2023}
Tanmaya Mishra et~al.
\newblock {Search for gravitational-wave bursts in the third Advanced LIGO-Virgo run with coherent WaveBurst enhanced by machine learning}.
\newblock {\em Phys. Rev. D}, 107(6):062002, 2023.

\bibitem{cwb_manual}
coherent waveburst manual, https://gwburst.gitlab.io.
\newblock Accessed: 2025-03-10.

\bibitem{cwb2011}
S.~Klimenko et~al.
\newblock Localization of gravitational wave sources with networks of advanced detectors.
\newblock {\em Phys. Rev. D}, 83:102001, May 2011.

\bibitem{cWB2019_widerlook}
F.~Salemi, E.~Milotti, G.~A. Prodi, G.~Vedovato, C.~Lazzaro, S.~Tiwari, S.~Vinciguerra, M.~Drago, and S.~Klimenko.
\newblock Wider look at the gravitational-wave transients from gwtc-1 using an unmodeled reconstruction method.
\newblock {\em Phys. Rev. D}, 100:042003, Aug 2019.

\bibitem{cwbWDM}
V.~Necula et~al.
\newblock {Transient analysis with fast Wilson-Daubechies time-frequency transform}.
\newblock {\em J. Phys. Conf. Ser.}, 363:012032, 2012.

\bibitem{Mishra:2021tmu}
Tanmaya Mishra et~al.
\newblock {Optimization of model independent gravitational wave search for binary black hole mergers using machine learning}.
\newblock {\em Phys. Rev. D}, 104(2):023014, 2021.

\bibitem{XGBoost}
extreme gradient boosting, https://github.com/dmlc/xgboost.
\newblock Accessed: 2025-03-10.

\bibitem{Cabero:2019orq}
M.~Cabero et~al.
\newblock Blip glitches in advanced ligo data.
\newblock {\em Classical and Quantum Gravity}, 36:155010, 2019.

\bibitem{OpenData_2023}
R.~Abbott et~al.
\newblock Open data from the third observing run of ligo, virgo, kagra, and geo.
\newblock {\em The Astrophysical Journal Supplement Series}, 267(2):29, jul 2023.

\bibitem{LIGO-O3}
A.~Buikema, C.~Cahillane, G.~L. Mansell, C.~D. Blair, R.~Abbott, C.~Adams, R.~X. Adhikari, A.~Ananyeva, S.~Appert, K.~Arai, J.~S. Areeda, Y.~Asali, S.~M. Aston, C.~Austin, A.~M. Baer, M.~Ball, S.~W. Ballmer, S.~Banagiri, D.~Barker, L.~Barsotti, J.~Bartlett, B.~K. Berger, J.~Betzwieser, D.~Bhattacharjee, G.~Billingsley, S.~Biscans, R.~M. Blair, N.~Bode, P.~Booker, R.~Bork, A.~Bramley, A.~F. Brooks, D.~D. Brown, K.~C. Cannon, X.~Chen, A.~A. Ciobanu, F.~Clara, S.~J. Cooper, K.~R. Corley, S.~T. Countryman, P.~B. Covas, D.~C. Coyne, L.~E.~H. Datrier, D.~Davis, C.~Di~Fronzo, K.~L. Dooley, J.~C. Driggers, P.~Dupej, S.~E. Dwyer, A.~Effler, T.~Etzel, M.~Evans, T.~M. Evans, J.~Feicht, A.~Fernandez-Galiana, P.~Fritschel, V.~V. Frolov, P.~Fulda, M.~Fyffe, J.~A. Giaime, K.~D. Giardina, P.~Godwin, E.~Goetz, S.~Gras, C.~Gray, R.~Gray, A.~C. Green, E.~K. Gustafson, R.~Gustafson, J.~Hanks, J.~Hanson, T.~Hardwick, R.~K. Hasskew, M.~C. Heintze, A.~F. Helmling-Cornell, N.~A. Holland, J.~D. Jones, S.~Kandhasamy, S.~Karki, M.~Kasprzack, K.~Kawabe, N.~Kijbunchoo, P.~J. King, J.~S. Kissel, Rahul Kumar, M.~Landry, B.~B. Lane, B.~Lantz, M.~Laxen, Y.~K. Lecoeuche, J.~Leviton, J.~Liu, M.~Lormand, A.~P. Lundgren, R.~Macas, M.~MacInnis, D.~M. Macleod, S.~M\'arka, Z.~M\'arka, D.~V. Martynov, K.~Mason, T.~J. Massinger, F.~Matichard, N.~Mavalvala, R.~McCarthy, D.~E. McClelland, S.~McCormick, L.~McCuller, J.~McIver, T.~McRae, G.~Mendell, K.~Merfeld, E.~L. Merilh, F.~Meylahn, T.~Mistry, R.~Mittleman, G.~Moreno, C.~M. Mow-Lowry, S.~Mozzon, A.~Mullavey, T.~J.~N. Nelson, P.~Nguyen, L.~K. Nuttall, J.~Oberling, Richard~J. Oram, B.~O'Reilly, C.~Osthelder, D.~J. Ottaway, H.~Overmier, J.~R. Palamos, W.~Parker, E.~Payne, A.~Pele, R.~Penhorwood, C.~J. Perez, M.~Pirello, H.~Radkins, K.~E. Ramirez, J.~W. Richardson, K.~Riles, N.~A. Robertson, J.~G. Rollins, C.~L. Romel, J.~H. Romie, M.~P. Ross, K.~Ryan, T.~Sadecki, E.~J. Sanchez, L.~E. Sanchez, T.~R. Saravanan, R.~L. Savage, D.~Schaetzl, R.~Schnabel, R.~M.~S. Schofield, E.~Schwartz, D.~Sellers, T.~Shaffer, D.~Sigg, B.~J.~J. Slagmolen, J.~R. Smith, S.~Soni, B.~Sorazu, A.~P. Spencer, K.~A. Strain, L.~Sun, M.~J. Szczepa\ifmmode~\acute{n}\else \'{n}\fi{}czyk, M.~Thomas, P.~Thomas, K.~A. Thorne, K.~Toland, C.~I. Torrie, G.~Traylor, M.~Tse, A.~L. Urban, G.~Vajente, G.~Valdes, D.~C. Vander-Hyde, P.~J. Veitch, K.~Venkateswara, G.~Venugopalan, A.~D. Viets, T.~Vo, C.~Vorvick, M.~Wade, R.~L. Ward, J.~Warner, B.~Weaver, R.~Weiss, C.~Whittle, B.~Willke, C.~C. Wipf, L.~Xiao, H.~Yamamoto, Hang Yu, Haocun Yu, L.~Zhang, M.~E. Zucker, and J.~Zweizig.
\newblock Sensitivity and performance of the advanced ligo detectors in the third observing run.
\newblock {\em Phys. Rev. D}, 102:062003, Sep 2020.

\bibitem{VirgoDetCharO3}
F~Acernese, M~Agathos, A~Ain, S~Albanesi, A~Allocca, A~Amato, T~Andrade, N~Andres, M~Andrés-Carcasona, T~Andrić, S~Ansoldi, S~Antier, T~Apostolatos, E~Z Appavuravther, M~Arène, N~Arnaud, M~Assiduo, S~Assis~de Souza~Melo, P~Astone, F~Aubin, S~Babak, F~Badaracco, M~K M~Bader, S~Bagnasco, J~Baird, T~Baka, G~Ballardin, G~Baltus, B~Banerjee, C~Barbieri, P~Barneo, F~Barone, M~Barsuglia, D~Barta, A~Basti, M~Bawaj, M~Bazzan, F~Beirnaert, M~Bejger, I~Belahcene, V~Benedetto, M~Berbel, S~Bernuzzi, D~Bersanetti, A~Bertolini, U~Bhardwaj, A~Bianchi, S~Bini, M~Bischi, M~Bitossi, M-A Bizouard, F~Bobba, M~Boër, G~Bogaert, M~Boldrini, L~D Bonavena, F~Bondu, R~Bonnand, B~A Boom, V~Boschi, V~Boudart, Y~Bouffanais, A~Bozzi, C~Bradaschia, M~Branchesi, M~Breschi, T~Briant, A~Brillet, J~Brooks, G~Bruno, F~Bucci, T~Bulik, H~J Bulten, D~Buskulic, C~Buy, G~S Cabourn~Davies, G~Cabras, R~Cabrita, G~Cagnoli, E~Calloni, M~Canepa, S~Canevarolo, M~Cannavacciuolo, E~Capocasa, G~Carapella, F~Carbognani, M~Carpinelli, G~Carullo, J~Casanueva~Diaz, C~Casentini, S~Caudill, F~Cavalier, R~Cavalieri, G~Cella, P~Cerdá-Durán, E~Cesarini, W~Chaibi, P~Chanial, E~Chassande-Mottin, S~Chaty, F~Chiadini, G~Chiarini, R~Chierici, A~Chincarini, M~L Chiofalo, A~Chiummo, S~Choudhary, N~Christensen, G~Ciani, P~Ciecielag, M~Cieślar, M~Cifaldi, R~Ciolfi, F~Cipriano, S~Clesse, F~Cleva, E~Coccia, E~Codazzo, P-F Cohadon, D~E Cohen, A~Colombo, M~Colpi, L~Conti, I~Cordero-Carrión, S~Corezzi, D~Corre, S~Cortese, J-P Coulon, M~Croquette, J~R Cudell, E~Cuoco, M~Curyło, P~Dabadie, T~Dal~Canton, S~Dall’Osso, G~Dálya, B~D’Angelo, S~Danilishin, S~D’Antonio, V~Dattilo, M~Davier, D~Davis, J~Degallaix, M~De~Laurentis, S~Deléglise, F~De~Lillo, D~Dell’Aquila, W~Del~Pozzo, F~De~Matteis, A~Depasse, R~De~Pietri, R~De~Rosa, C~De~Rossi, R~De~Simone, L~Di~Fiore, C~Di~Giorgio, F~Di~Giovanni, M~Di~Giovanni, T~Di~Girolamo, A~Di~Lieto, A~Di~Michele, S~Di~Pace, I~Di~Palma, F~Di~Renzo, L~D’Onofrio, M~Drago, J-G Ducoin, U~Dupletsa, O~Durante, D~D’Urso, P-A Duverne, M~Eisenmann, L~Errico, D~Estevez, F~Fabrizi, F~Faedi, V~Fafone, S~Farinon, G~Favaro, M~Fays, E~Fenyvesi, I~Ferrante, F~Fidecaro, P~Figura, A~Fiori, I~Fiori, R~Fittipaldi, V~Fiumara, R~Flaminio, J~A Font, S~Frasca, F~Frasconi, A~Freise, O~Freitas, G~G Fronzé, B~U Gadre, R~Gamba, B~Garaventa, F~Garufi, G~Gemme, A~Gennai, Archisman Ghosh, B~Giacomazzo, L~Giacoppo, P~Giri, F~Gissi, S~Gkaitatzis, B~Goncharov, M~Gosselin, R~Gouaty, A~Grado, M~Granata, V~Granata, G~Greco, G~Grignani, A~Grimaldi, S~J Grimm, P~Gruning, D~Guerra, G~M Guidi, G~Guixé, Y~Guo, P~Gupta, L~Haegel, O~Halim, O~Hannuksela, T~Harder, K~Haris, J~Harms, B~Haskell, A~Heidmann, H~Heitmann, P~Hello, G~Hemming, E~Hennes, S~Hild, D~Hofman, V~Hui, B~Idzkowski, A~Iess, P~Iosif, T~Jacqmin, P-E Jacquet, S~P Jadhav, J~Janquart, K~Janssens, P~Jaranowski, V~Juste, C~Kalaghatgi, C~Karathanasis, S~Katsanevas, F~Kéfélian, N~Khetan, G~Koekoek, S~Koley, M~Kolstein, A~Królak, P~Kuijer, P~Lagabbe, D~Laghi, M~Lalleman, A~Lamberts, I~La~Rosa, A~Lartaux-Vollard, C~Lazzaro, P~Leaci, A~Lemaître, M~Lenti, E~Leonova, N~Leroy, N~Letendre, K~Leyde, F~Linde, L~London, A~Longo, M~Lopez~Portilla, M~Lorenzini, V~Loriette, G~Losurdo, D~Lumaca, A~Macquet, C~Magazzù, M~Magnozzi, E~Majorana, I~Maksimovic, N~Man, V~Mangano, M~Mantovani, M~Mapelli, F~Marchesoni, D~Marín~Pina, F~Marion, A~Marquina, S~Marsat, F~Martelli, M~Martinez, V~Martinez, A~Masserot, S~Mastrogiovanni, Q~Meijer, A~Menendez-Vazquez, L~Mereni, M~Merzougui, A~Miani, C~Michel, L~Milano, A~Miller, B~Miller, E~Milotti, Y~Minenkov, Ll~M~Mir, M~Miravet-Tenés, M~Montani, F~Morawski, B~Mours, C~M Mow-Lowry, S~Mozzon, F~Muciaccia, Suvodip Mukherjee, R~Musenich, A~Nagar, V~Napolano, I~Nardecchia, H~Narola, L~Naticchioni, J~Neilson, C~Nguyen, S~Nissanke, E~Nitoglia, F~Nocera, G~Oganesyan, C~Olivetto, G~Pagano, G~Pagliaroli, C~Palomba, P~T H~Pang, F~Pannarale, F~Paoletti, A~Paoli, A~Paolone, G~Pappas, D~Pascucci, A~Pasqualetti, R~Passaquieti, D~Passuello, B~Patricelli, R~Pedurand, M~Pegoraro, A~Perego, A~Pereira, C~Périgois, A~Perreca, S~Perriès, D~Pesios, K~S Phukon, O~J Piccinni, M~Pichot, M~Piendibene, F~Piergiovanni, L~Pierini, V~Pierro, G~Pillant, M~Pillas, F~Pilo, L~Pinard, I~M Pinto, M~Pinto, K~Piotrzkowski, A~Placidi, E~Placidi, W~Plastino, R~Poggiani, E~Polini, E~K Porter, R~Poulton, M~Pracchia, T~Pradier, M~Principe, G~A Prodi, P~Prosposito, A~Puecher, M~Punturo, F~Puosi, P~Puppo, G~Raaijmakers, N~Radulesco, P~Rapagnani, M~Razzano, T~Regimbau, L~Rei, P~Rettegno, B~Revenu, A~Reza, F~Ricci, G~Riemenschneider, S~Rinaldi, F~Robinet, A~Rocchi, L~Rolland, M~Romanelli, R~Romano, A~Romero, S~Ronchini, L~Rosa, D~Rosińska, S~Roy, D~Rozza, P~Ruggi, J~Sadiq, O~S Salafia, L~Salconi, F~Salemi, A~Samajdar, N~Sanchis-Gual, A~Sanuy, B~Sassolas, S~Sayah, S~Schmidt, M~Seglar-Arroyo, D~Sentenac, V~Sequino, Y~Setyawati, A~Sharma, N~S Shcheblanov, M~Sieniawska, L~Silenzi, N~Singh, A~Singha, V~Sipala, J~Soldateschi, K~Soni, V~Sordini, F~Sorrentino, N~Sorrentino, R~Soulard, V~Spagnuolo, M~Spera, P~Spinicelli, C~Stachie, D~A Steer, J~Steinlechner, S~Steinlechner, N~Stergioulas, G~Stratta, M~Suchenek, A~Sur, B~L Swinkels, P~Szewczyk, M~Tacca, A~J Tanasijczuk, E~N Tapia San~Martín, C~Taranto, A~E Tolley, M~Tonelli, A~Torres-Forné, I~Tosta~e Melo, A~Trapananti, F~Travasso, M~Trevor, M~C Tringali, L~Troiano, A~Trovato, L~Trozzo, K~W Tsang, K~Turbang, M~Turconi, A~Utina, M~Valentini, N~van Bakel, M~van Beuzekom, M~van Dael, J~F J~van~den Brand, C~Van Den~Broeck, H~van Haevermaet, J~V van Heijningen, N~van Remortel, M~Vardaro, M~Vasúth, G~Vedovato, D~Verkindt, P~Verma, F~Vetrano, A~Viceré, V~Villa-Ortega, J-Y Vinet, A~Virtuoso, H~Vocca, R~C Walet, M~Was, A~R Williamson, J~L Willis, A~Zadrożny, T~Zelenova, and J-P Zendri.
\newblock Virgo detector characterization and data quality: results from the o3 run.
\newblock {\em Classical and Quantum Gravity}, 40(18):185006, aug 2023.

\bibitem{GW150914minimalassumptions}
B.~P. Abbott et~al.
\newblock Observing gravitational-wave transient gw150914 with minimal assumptions.
\newblock {\em Phys. Rev. D}, 93:122004, Jun 2016.

\bibitem{RateAndPopO3}
R.~Abbott and other.
\newblock Population of merging compact binaries inferred using gravitational waves through gwtc-3.
\newblock {\em Phys. Rev. X}, 13:011048, Mar 2023.

\bibitem{Was2010}
Michał Wąs, Marie-Anne Bizouard, Violette Brisson, Fabien Cavalier, Michel Davier, Patrice Hello, Nicolas Leroy, Florent Robinet, and Miltiadis Vavoulidis.
\newblock Limitations of the time slide method of background estimation.
\newblock {\em Classical and Quantum Gravity}, 27(19):194014, sep 2010.

\bibitem{Mishra:2024zzs}
Tanmaya Mishra et~al.
\newblock {Gravitational Waves Detected by a Burst Search in LIGO/Virgo's Third Observing Run}.
\newblock 10 2024.

\bibitem{Bini:2023gaj}
Sophie Bini et~al.
\newblock {Search for hyperbolic encounters of compact objects in the third LIGO-Virgo-KAGRA observing run}.
\newblock {\em Phys. Rev. D}, 109(4):042009, 2024.

\bibitem{Xu:2023ioz}
Yumeng Xu, Shubhanshu Tiwari, and Marco Drago.
\newblock {PycWB: A User-friendly, Modular, and Python-based Framework for Gravitational Wave Unmodelled Search}.
\newblock 8 2023.

\bibitem{pycwb_repo}
\url{https://github.com/PycWB/pycwb}.

\bibitem{V.Tiwari_2016}
V~Tiwari, S~Klimenko, V~Necula, and G~Mitselmakher.
\newblock Reconstruction of chirp mass in searches for gravitational wave transients.
\newblock {\em Classical and Quantum Gravity}, 33(1):01LT01, dec 2015.

\end{thebibliography}

\end{document}